\documentclass[fleqn, usenatbib]{mn2e}

\usepackage{fixltx2e} 

\usepackage{graphicx}

\usepackage{amsmath}
\usepackage{amssymb}
\usepackage{amsfonts}
\usepackage{mathptmx}

\usepackage{natbib}
\usepackage{aas_macros}

\title [Galaxy clustering as a function of stellar mass]
{A new methodology to test galaxy formation models using the dependence of clustering on stellar mass}

\author[David~J.~R.~Campbell et al.]
{David~J.~R.~Campbell,$^1$\thanks{Email: d.j.r.campbell@durham.ac.uk}
Carlton~M.~Baugh,$^1$ Peter~D.~Mitchell,$^1$ John~C.~Helly,$^1$
\newauthor Violeta~Gonzalez-Perez,$^1$ Cedric~G.~Lacey,$^1$ Claudia~del~P.~Lagos,$^2$ Vimal Simha,$^1$
\newauthor Daniel~J.~Farrow$^{1,3}$
\vspace{2mm}
\\
$^1$Institute for Computational Cosmology,
Department of Physics, Durham University, South Road, Durham, DH1 3LE,
UK.
\\
$^2$European Southern Observatory, Karl-Schwarzschild-Strasse 2, 85748, Garching, Germany.
\\
$^3$Max-Planck-Institut fuer extraterrestrische Physik, Giessenbachstrasse 1, 85748 Garching, Germany.
}

\begin{document}

\label{firstpage}
\date{Accepted 2015 June 10. Received 2015 May 13; in original form 2014 December 11}
\pagerange{\pageref{firstpage}--\pageref{lastpage}}
\pubyear{2015}

\maketitle

\begin{abstract}
We present predictions for the two-point correlation function of galaxy clustering as a function of stellar mass, computed using two new versions of the \textsc{galform} semi-analytic galaxy formation model. These models make use of a high resolution, large volume N-body simulation, set in the WMAP7 cosmology. One model uses a universal stellar initial mass function (IMF), while the other assumes different IMFs for quiescent star formation and bursts. Particular consideration is given to how the assumptions required to estimate the stellar masses of observed galaxies (such as the choice of IMF, stellar population synthesis model and dust extinction) influence the perceived dependence of galaxy clustering on stellar mass. Broad-band spectral energy distribution fitting is carried out to estimate stellar masses for the model galaxies in the same manner as in observational studies. We show clear differences between the clustering signals computed using the true and estimated model stellar masses. As such, we highlight the importance of applying our methodology to compare theoretical models to observations. We introduce an alternative scheme for the calculation of the merger timescales for satellite galaxies in \textsc{galform}, which takes into account the dark matter subhalo information from the simulation. This reduces the amplitude of small-scale clustering. The new merger scheme offers improved or similar agreement with observational clustering measurements, over the redshift range $0<z<0.7$. We find reasonable agreement with clustering measurements from GAMA, but find larger discrepancies for some stellar mass ranges and separation scales with respect to measurements from SDSS and VIPERS, depending on the \textsc{galform} model used.
\end{abstract}


\section{Introduction}

Galaxy formation involves the interplay between a variety of physical processes, such as the suppression of star formation in massive haloes due to the shutting down of gas cooling by active galactic nuclei, and the heating of the interstellar medium by supernovae. These processes vary in importance with redshift and the mass of the dark matter halo which hosts the galaxy. Semi-analytic models use simplified expressions and approximations to describe these complex physical processes, allowing predictions to be generated for how dark matter haloes are populated with galaxies \citep{Benson_2000_first,Cole_2000,Hatton_2003,Bower_2006,Croton_2006,Monaco_2007,Somerville_2008}.

The model processes are constrained through comparison to observed statistics of the galaxy population, such as the luminosity function (e.g.\ \citealp{Gonzalez_2014}). We can also constrain the galaxy formation physics by measuring galaxy clustering as a function of intrinsic properties, such as stellar mass. Appealing to the clustering as a function of galaxy properties provides a constraint on how dark matter haloes are populated with galaxies \citep{Kauffmann_1999_II,Benson_2000_second}. Galaxy formation models have been used to make predictions for comparison to observational measurements of the clustering as a function of different properties, such as luminosity, star formation rate, and stellar mass (e.g.\ \citealp{Norberg_2001,Kim_2009,Meneux_2009,Guo_2011,Li_2012,Guo_2013,Henriques_2013,Marulli_2013,Bielby_2014}). Some models explicitly consider galaxy clustering when setting their parameters, e.g.\ \cite{Guo_2011}. For models which have been constructed without using galaxy clustering as a constraint, comparison to clustering measurements is a test of their predictive power. The models considered in this paper have not been calibrated to reproduce any clustering measurements, or stellar mass function data.

Despite the perceived complexity of semi-analytic models, \cite{Contreras_2013} show that the clustering predictions made by different groups are robust, at least for samples defined by stellar mass. They considered the Durham \textsc{galform} \citep{Bower_2006,Font_2008} and Munich \textsc{lgalaxies} \citep{Bertone_2007,deLucia_Blaizot_2007,Guo_2011} model families. These two sets of models differ in their implementation of various physical processes of galaxy formation. \citeauthor{Contreras_2013} find differences in the clustering predictions on small scales, which they trace to the handling of galaxy mergers. We explore the impact of the choice of galaxy merger scheme further in this paper.

The stellar masses of real galaxies have to be derived from observable properties such as broad-band photometry, which requires the assumption of a star formation history (SFH), stellar initial mass function (IMF), stellar population synthesis (SPS) model, and dust extinction model (e.g.\ \citealp{Pforr_2012,Mitchell_2013}). The choice of the particular set of assumptions used has an impact on the recovered masses, and so there are important systematic uncertainties inherent in the derived stellar masses of observed galaxies. These uncertainties may influence comparisons of galaxy clustering as a function of stellar mass, between theoretical models and observational measurements. Thus if the clustering as a function of stellar mass is to be used to constrain theoretical models, we must take care to treat appropriately the difficulties in estimating the stellar masses of observed galaxies.

\cite{Contreras_2013} compared theoretical predictions by examining galaxy samples of fixed abundance. We extend this study, motivated by the work of \citet[see also \citealp{Marchesini_2009} and \citealp{Pforr_2012}]{Mitchell_2013}, and demonstrate a new methodology which we find is essential for comparing model predictions to observational clustering data. Our technique consists of carrying out broad-band spectral energy distribution (SED) fitting to compute stellar mass estimates for model galaxies, in the manner that is typically applied to observed galaxies. Such a treatment is particularly important for understanding the influence of dust extinction, and the impact of Eddington bias in higher mass bins where the stellar mass function is steep.

We present the clustering predictions of two new \textsc{galform} variants: the models of \cite{Gonzalez_2014} and Lacey et al.\ (in preparation), hereafter Gon14 and Lac14. These models have been calibrated to reproduce observations in the WMAP7 cosmology \citep{Komatsu_2011}. The models make use of the MS-W7 N-body simulation \citep{Guo_2013}, which is a new version of the Millennium Simulation \citep{Springel_2005} updated to use the WMAP7 cosmological parameters. These models take advantage of extensions to the galaxy formation physics implemented in \textsc{galform}, making use of an empirical law to determine star formation rates \citep{Lagos_2011b}. For discussions of the implications for galaxy formation models of the WMAP7 cosmology, see \cite{Guo_2013} and Gon14.

We compare the model predictions to derived results from the Sloan Digital Sky Survey (SDSS; \citealp{York_2000,Abazajian_2009}), the Galaxy and Mass Assembly Survey (GAMA; \citealp{Driver_2011}), and the VIMOS Public Extragalactic Redshift Survey (VIPERS; \citealp{Guzzo_2014}). Together, these surveys have measured the clustering of galaxies as a function of stellar mass up to redshift $z\sim1$ (\citealp{Li_2006}; Farrow et al.\ in preparation; \citealp{Marulli_2013}). In particular, we compare to these observational results at redshifts 0.1, 0.2, and 0.6 respectively. The specific star formation rate falls dramatically over this interval (e.g.\ \citealp{Weinmann_2011}), so the interplay between the different galaxy formation processes could change over the cosmic times considered. Through comparing to these measurements, we assess both the importance of carrying out SED fitting to the model photometry, and the level of agreement between the measurements and the model predictions.

The layout of this paper is as follows.
Details of the galaxy formation models and observational data are given in \S\ref{sec:gf} and \S\ref{sec:obsdata}.
Our methodology is described in \S\ref{sec:analysis}.
Our results are presented and discussed in \S\ref{sec:results}.
Concluding remarks are given in \S\ref{sec:conclusions}.

All magnitudes are on the AB system \citep{Oke_1974}.
The assumed $\Lambda$CDM cosmological parameters are listed in \S\ref{sec:gf_simulation}.
Comoving length units are used throughout this paper.

\section{Theoretical Modelling}
\label{sec:gf}

This section describes the galaxy formation models used in our study of galaxy clustering as a function of stellar mass. We first give an overview of \textsc{galform} (\S\ref{sec:gf_overview}), followed by a description of the N-body simulation used (\S\ref{sec:gf_simulation}), then contrast the two models compared (\S\ref{sec:gf_models}), and describe a new scheme for the treatment of satellite galaxy orbits and merger time-scales (\S\ref{sec:gf_sdf}).

\subsection{Overview of \textsc{galform}}
\label{sec:gf_overview}

Galaxy formation within dark matter haloes, as followed in recent \textsc{galform} variants, can be broken down into several key processes:
(i) formation and hierarchical growth of dark matter haloes,
(ii) shock-heating of baryonic material falling into haloes, followed by radiative cooling and disc formation,
(iii) quiescent star formation in discs (and bursts due to instabilities),
(iv) suppression of gas cooling (and hence of star formation) through feedback from supernovae, active galactic nuclei (AGN), and photo-ionization of the intergalactic medium,
(v) chemical enrichment of the stars and gas,
and (vi) mergers of satellite galaxies with the central galaxy of their halo, due to dynamical friction, which can cause bursts of star formation.
For an overview of the development of the \textsc{galform} model, see \cite{Benson_Bower_2010}. For details of the galaxy formation physics, see \cite{Baugh_2006} and \cite{Benson_2010}.

In order to connect the predictions of galaxy formation models to the properties of observed galaxies, a stellar population synthesis (SPS) model must be assumed, along with a model to describe dust attenuation. SPS models, such as those of \cite{Bruzual_Charlot_2003}, \cite{Maraston_2005}, and \cite{Conroy_2009}, compute the spectral energy distribution (SED) of a coeval stellar population with a given initial metallicity, as a function of age. Convolving this stellar SED with the star formation history of a galaxy (i.e.\ its star formation rate as a function of time; SFH), taking into account its chemical enrichment history (the metallicity of stars forming at a given time), yields the SED of the galaxy itself \citep{Cole_2000}. SPS models require a stellar initial mass function (IMF) to be specified, which gives the distribution of masses of stars formed in a given episode of star formation.

Attenuation of starlight by dust in \textsc{galform} is modelled in a physically motivated way, in which the stars and dust are mixed together, based on radiative transfer and the geometry of a disc and bulge \citep{Silva_1998,Ferrara_1999,Cole_2000,Lacey_2011,Gonzalez_2013_2}. For a given photometric band, the galactic SED is weighted by the wavelength response of the filter and integrated to yield the flux.

\subsection{The MS-W7 simulation}
\label{sec:gf_simulation}

The cosmological parameters from WMAP7 \citep{Komatsu_2011} have been used in an N-body simulation similar to the Millennium Simulation of \cite{Springel_2005}. This MS-W7 simulation \citep{Guo_2013} has present day density parameters of $\Omega_\mathrm{m}=0.272$, $\Omega_\mathrm{b}=0.0455$, and $\Omega_\Lambda=0.728$, for matter, baryons, and dark energy, respectively. The present day Hubble parameter is $H=100~h~\mathrm{km\,s^{-1}\,Mpc^{-1}}$, where $h=0.704$. The spectral index of primordial fluctuations is $n_\mathrm{s}=0.967$, and the linear perturbation amplitude is $\sigma_8=0.810$. The simulation follows $2160^3$ particles from redshift 127, in a volume of comoving side $L_\mathrm{box} = 500~h^{-1}\,\mathrm{Mpc}$.

\begin{table*}
\caption{
Observed galaxy samples.
The galaxy numbers refer to galaxies contributing to the correlation function $w_\mathrm{p}(\sigma)$ in Figs.\ \ref{fig:wp_SDSS}, \ref{fig:wp_GAMA}, and \ref{fig:wp_VIPERS}.
$r$ and $i$ are the SDSS $r$-band and CFHTLS $i$-band apparent magnitudes, respectively.
$M_{0.1_r}$ is the SDSS $r$-band absolute magnitude (at $z=0.1$).
The median redshifts for GAMA and VIPERS are given in order of increasing stellar mass interval (see Figs.\ \ref{fig:wp_GAMA} and \ref{fig:wp_VIPERS}).
The \textsc{galform} comparison redshifts are restricted to the set of output snapshots of the MS-W7 simulation.
$\pi_\mathrm{max}$ is the $w_\mathrm{p}(\sigma)$ integration limit (see Eqn.\ \ref{eq:wp_def}).
The lower part of the table lists the \protect\citet{Schechter_1976} function parameters $\alpha$, $\tilde{M}_\star$, and $\tilde{\Phi}$ (see Eqn.\ \ref{eq:schfn}) for the stellar mass functions of the samples.
The SDSS mass function fit is as measured by \protect\citet{Wang_2006}.
The GAMA stellar mass function has been measured by \protect\citet{Baldry_2012} for $z<0.06$ (see \S\ref{sec:analysis_sm_schfun}), and is represented by the sum of two \protect\citeauthor{Schechter_1976} functions (left, right), with a common characteristic mass.
The VIPERS fit is for $0.5<z<0.6$ \protect\citep{Davidzon_2013}.
}
\centering
\begin{tabular}{ccccc}
\hline
Sample property                 &SDSS		&GAMA		&VIPERS \\
\hline
Source                                  &\cite{Li_2006}
                                                                &Farrow et al.\ (in preparation)
                                                                &\cite{Marulli_2013} \\
Galaxies used           &157884
                                                                &50770
                                                                &17100 \\
Apparent magnitude limits
                                                                &$14.5 < r < 17.77$
                                                                &$r<19.8$
                                                        &$ i<22.5$ \\
Absolute magnitude limits
                                                        &$-23<M_{0.1_r} - 5\log_{10}(h) <-16$
                                                        &
                                                        & \\
Redshift range              &$0.01< z <0.3$
                                                                &$0.14< z < 0.24$
                                                                &$0.5< z <0.7$ \\
Median redshift                 &$\sim0.1$
                                                                &0.19, 0.20
                                                                &0.62, 0.62, 0.62 \\
\textsc{galform} redshift for comparison
                                                   &0.089
                                                                &0.17
                                                                &0.62 \\
$\pi_\mathrm{max}~[h^{-1}\,\mathrm{Mpc}]$
                                                                &40
                                                                &47
                                                                &30 \\
\hline
$\alpha$					&$-1.073 \pm 0.003$
                                                                &$-0.35\pm0.18, -1.47\pm0.05$
                                                                &$-0.95^{+0.03}_{-0.02}$ \\
$\log_{10}(\tilde{M}_\star~[h^{-2}\,\mathrm{M}_\odot])$
                                        &$10 + \log_{10}(4.11\pm0.02)$
                                        &$(10.66\pm0.05) + \log_{10}(0.7^2)$
                                        & $10.87^{+0.02}_{-0.02} + \log_{10}(0.7^2)$ \\
$\tilde{\Phi}~[10^{-3}\,h^3\,\mathrm{Mpc}^{-3}\,\mathrm{dex}^{-1}]$
                                        &$(20.4 \pm 0.1)/\ln(10)$
                                        &$(3.96\pm0.34, 0.79\pm0.23) / 0.7^{3}$
                                        &$1.42^{+0.06}_{-0.07} / 0.7^{3}$	\\
\hline
\end{tabular}
\label{tab:obspapers}
\end{table*}

\subsection{The Gon14 and Lac14 models}
\label{sec:gf_models}

The Gon14 and Lac14 \textsc{galform} models are based on the halo merger trees of the MS-W7 simulation. They are separate developments of the \cite{Lagos_2012} model, which used an empirical star formation rate law introduced by  \cite{Lagos_2011b}. The \cite{Lagos_2012} model in turn is based on that of \cite{Bower_2006}, which introduced AGN feedback into \textsc{galform}. The new models will be made publicly available in the Millennium Database.\footnote{http://virgodb.cosma.dur.ac.uk:8080/Millennium}

The IMF $\zeta(M_\star)$ is defined such that the number of newly formed stars, per solar mass, with stellar mass, $M_\star$, between $\log_{10}(M_\star/M_\mathrm{ref})$ and $\log_{10}(M_\star/M_\mathrm{ref})+\mathrm{d}\log_{10}(M_\star/M_\mathrm{ref})$ is given by $\zeta (M_\star)\, \mathrm{d}\log_{10}(M_\star/M_\mathrm{ref})$, for some mass unit $M_\mathrm{ref}$. A power law is often used, such that,
\begin{equation}
\zeta (M_\star)
\, \mathrm{d}\log_{10}\left(\frac{M_\star}{M_\mathrm{ref}}\right)
= \zeta_0 \,\left(\frac{M_\star}{M_\mathrm{ref}}\right)^{-x}
\, \mathrm{d}\log_{10}\left(\frac{M_\star}{M_\mathrm{ref}}\right)  ~,
\label{eq:imf}
\end{equation}
for some normalisation $\zeta_0$ and slope $x$. The Gon14 model uses a \cite{Kennicutt_1983} IMF, which is a broken power law. This has a slope of $x=0.4$ for $M_\star<1~\mathrm{M}_\odot$, and $x=1.5$ for $M_\star>1~\mathrm{M}_\odot$. In the Lac14 model, distinct IMFs are employed in quiescent star formation and bursts. In the former case, a \citeauthor{Kennicutt_1983} IMF is used as in Gon14. However, for bursts, the IMF is taken to be a single power law with $x=1$. Such non-universality of the IMF is argued to be necessary to match the observed number counts and redshift distribution of sub-millimetre galaxies \citep{Baugh_2005}.

The SPS model used in the Gon14 model is a private release of the \citeauthor{Bruzual_Charlot_2003} series from 1999, which is intermediate between \cite{Bruzual_Charlot_1993} and \cite{Bruzual_Charlot_2003}. The Lac14 model uses the \cite{Maraston_2005} SPS model. The \citeauthor{Maraston_2005} model attributes much more luminosity to stars in the thermally pulsating asymptotic giant branch (TP-AGB) phase than is done in the \citeauthor{Bruzual_Charlot_2003} model. Such stars emit strongly in the near infra-red (NIR; e.g.\ \citealp{MacArthur_2010}). The influence of TP-AGB stars in the \citeauthor{Maraston_2005} model has been the source of some debate in the literature (e.g.\ \citealp{Marigo_2007}; \citealp{Kriek_2010}; \citealp{MacArthur_2010}; \citealp{Zibetti_2013}). Gon14 study the influence of using various alternative SPS models, including that of \citeauthor{Maraston_2005}. They find that the choice of SPS model does not affect the evolution of the rest-frame optical and ultra-violet (UV) luminosity functions, but models incorporating strong TP-AGB emission yield significantly different evolution of the rest-frame NIR luminosity function (see also \citealp{Tonini_2009} and \citealp{Henriques_2011}). This choice gives an improved match to the bright end of the rest-frame $K$-band luminosity function at high redshifts in the Lac14 model.

Additional parameter differences between the Gon14 and Lac14 models are described in Appendix \ref{sec:params}.

\subsection{Subhalo dynamical friction for satellites}
\label{sec:gf_sdf}

In the standard models considered here, time-scales for the merging of satellites with the central galaxy in their host halo due to dynamical friction are computed in the models as described by \cite{Cole_2000}. This method assumes that when a new halo forms, each satellite galaxy enters the halo on a random orbit. The merger time-scale is then computed using an analytical formula which assumes the halo to be isothermal. While the Gon14 model makes use of the equations presented by \cite{Lacey_Cole_1993}, a modified expression is used in the Lac14 model. This expression has been empirically fitted to numerical simulations to account for the tidal stripping of subhaloes \citep{Jiang_2008,Jiang_2010}, but otherwise the treatment is the same; i.e.\ an analytic time-scale is computed as soon as a galaxy enters a larger halo. The satellite is considered to have merged with its central galaxy once the merger time-scale has elapsed, provided that this transpires before the halo merges to form a larger system, in which case a new merger time-scale is computed. Note that this scheme does not take into account that the satellite may still reside in a resolvable dark matter subhalo at the time the merger takes place.

We have implemented an alternative treatment of mergers, which makes use of the subhalo information from the simulation. The motivation for using this new scheme is that it is more faithful to the underlying N-body simulation, minimising the reliance on analytically determined orbits. Instead of assuming random initial orbits for satellites, they track the positions of their associated subhaloes. For cases where the subhalo containing a satellite can no longer be resolved following mass stripping, the position and velocity of the subhalo when it was last identified are used to compute a merger time-scale. This time-scale is then used in the same way as in the default scheme described above. The merger time-scale calculation assumes an NFW \citep{Navarro_1996} halo mass distribution to compute the orbital parameters of the satellite, combined with the analytical time-scale result of \cite{Lacey_Cole_1993}. If a halo formation event occurs at a time after the subhalo is lost, a new merger time-scale for the satellite is calculated in the same way, using instead the position and velocity of the particle which was the most bound particle of the subhalo when it was last identified.

In this new \textsc{galform} merger scheme, a satellite galaxy is not allowed to merge while it remains associated with a resolved subhalo. This treatment is similar to the scheme employed in the \textsc{lgalaxies} model. The choice of merger scheme has an impact on the small-scale clustering, and contributed to the differences between the predictions of \textsc{lgalaxies} and \textsc{galform} reported by \cite{Contreras_2013}. The differences between the clustering predictions using the two types of merger scheme can be explained in terms of the concentration of the radial distribution of satellites, and also changes in the number of satellites. Requiring satellites to track their resolved subhaloes, rather than computing an analytical merger time-scale as soon as a galaxy becomes a satellite, results in a more radially extended distribution of satellites (as demonstrated by \citeauthor{Contreras_2013}).

No model parameters have been recalibrated when using the new merger scheme. This would be likely to involve changing the strength of AGN feedback, and the timescale for gas return from supernovae. We leave such calibration for a future paper. When comparing the model predictions using the new merger scheme to observational estimates of galaxy clustering  as a function of stellar mass (in \S\ref{sec:results_comp_sdf}), we require that the model stellar mass functions reproduce those inferred from observations, through abundance matching (see \S\ref{sec:analysis_sm_schfun}). It is possible that making use of the subhalo mass at the time it was last identified results in shorter merger timescales than would be computed if any subsequent mass stripping of the subhalo could be taken into account.

\section{SED Fitting and Observational Data}
\label{sec:obsdata}

In theoretical models such as \textsc{galform}, the galactic stellar masses are predicted from the star formation histories. However, for observed galaxies, the stellar masses are not known directly but must be derived from observables. SED fitting is a popular technique for computing stellar masses. This section first describes the SED fitting procedure and then gives details of the observational data considered in this paper. We consider different surveys in order to probe a range of lookback times (see Table \ref{tab:obspapers}). A detailed discussion of SED fitting can be found in \cite{Mitchell_2013}.

\subsection{SED fitting}
\label{sec:obsdata_SED}

Broad-band SED fitting is essentially the reverse of the process described in \S\ref{sec:gf_overview} for computing galaxy SEDs and broad-band photometry in \textsc{galform}. A grid of template SEDs is generated, given an assumed SPS model, IMF, SFH (including assumptions about chemical enrichment), and dust extinction. The observed SED (i.e.\ broad-band photometry) is then used to identify the maximum-likelihood template SED (e.g.\ \citealp{Mitchell_2013}). The stellar mass associated with the best-fitting SED is then assigned to the observed galaxy. The SFH is usually taken to be of a simple exponentially declining form, in contrast to the complicated form predicted in theoretical galaxy formation models (see e.g.\ \citealp{Baugh_2006}). \citeauthor{Mitchell_2013} show that on average, the detailed form of the SFH is not important (see also \citealp{Simha_2014}). In SED fitting, dust attenuation is usually dealt with by assuming the so-called Calzetti law \citep{Calzetti_2000}, which is equivalent to assuming that the dust forms a uniform sheet between the galaxy and the observer. \textsc{galform} assumes a physically motivated distribution of dust in model galaxies (see \S\ref{sec:gf_overview}), and applies dust extinction in massive galaxies that is very different from the Calzetti law, resulting in systematic errors of up to an order of magnitude in $M_\star$ \citep{Mitchell_2013}. \cite{Conroy_2009} report that the uncertainties in stellar masses derived from broad-band SED fitting are in the region of 0.3 dex at redshift zero, considering the uncertainties in the details of different stages of stellar evolution, while at $z=2$ the uncertainty for bright red galaxies rises to $\sim0.6$ dex.

\subsection{Sloan Digital Sky Survey}
\label{sec:obsdata_SDSS}

The Sloan Digital Sky Survey (SDSS) uses photometry in the $u$, $g$, $r$, $i$, and $z$ bands to identify candidates for spectroscopic follow-up \citep{York_2000}. \cite{Li_2006} estimated the clustering of galaxies as a function of stellar mass using the New York University value-added galaxy catalogue \citep{Blanton_2005}, which is based on the second SDSS data release \citep{Abazajian_2004}. This catalogue has photometry covering 3514 square degrees, and spectroscopy covering 2627 square degrees (about 85 percent complete), for redshifts $z\lesssim0.3$. \citeauthor{Li_2006} define their sample of galaxies with the magnitude and redshift limits given in Table \ref{tab:obspapers}, yielding a total of 196238 galaxies. Subsets of this sample, defined in stellar mass, are used to study galaxy clustering.

\citeauthor{Li_2006} use the approach of \cite{Kauffmann_2003} to estimate stellar masses. The amplitude of the 4000~\AA\ break, $D_{4000}$, \citep{Balogh_1999} and the strength of the Balmer H-$\delta$ absorption line \citep{Worthey_1997} are measured from a spectrum obtained with a 3 arcsec diameter fibre. These measurements encode information about the age of the galaxy's stellar population ($D_{4000}$), and can be used as indicators of whether recent star formation has been predominantly quiescent, or due to bursts (H-$\delta$). The stellar mass-to-light ratio in the $z$-band is estimated for each galaxy, by fitting $D_{4000}$ and H-$\delta$ to a Monte Carlo library of stellar populations, based on the \cite{Bruzual_Charlot_2003} SPS model.
A \cite{Kroupa_2001} IMF is assumed in building the library, i.e.\
$x=-0.3$ for $M_\star<0.08~\mathrm{M}_\odot$,
$x=0.3$ for $0.08~\mathrm{M}_\odot<M_\star<0.5~\mathrm{M}_\odot$,
and $x=1.3$ for $M_\star>0.5~\mathrm{M}_\odot$ (see Eqn.\ \ref{eq:imf}).
Exponentially declining SFHs are used, with additional random bursts. The dust extinction applied in the best-fitting model is assumed for each galaxy, making use of a power law attenuation curve, corresponding to a foreground dust screen \citep{Charlot_2000}. The stellar mass is then found by using the derived mass-to-light ratio in the $z$-band, combined with full $z$-band photometry (i.e.\ not limited by the fibre diameter). In this way, the mass-to-light ratio and dust attenuation derived within 1.5 arcsec of the galactic centre are extrapolated over the full galaxy. The masses estimated following \citeauthor{Kauffmann_2003} using spectral features have been shown to have a scatter of about 0.1 dex with respect to those obtained using SED fitting to broad-band photometry \citep{Blanton_Roweis_2007,Li_White_2009}.

The clustering results presented by \citeauthor{Li_2006} use six bins spaced logarithmically in stellar mass,\footnote{
The stellar mass bin edges of \cite{Li_2006} have been converted from units of $\mathrm{M}_\odot$ to $h^{-2}\,\mathrm{M}_\odot$, using their $h=0.7$ \citep{Kauffmann_2003}.
}
covering $10^{8.69}<M_\star~[h^{-2}\,\mathrm{M}_\odot]<10^{11.69}$. All but the highest mass bin correspond to samples which are volume limited in stellar mass, where each volume limited stellar mass bin covers a different redshift interval. However, the highest mass bin is flux limited, and thus incomplete in stellar mass. A correction is made by \citeauthor{Li_2006} to the correlation function computed for this bin, by weighting the contribution from each galaxy pair by the maximum volume over which they could be detected in the survey volume. They find that applying the same approach to flux limited samples for the lower stellar masses produces good agreement with the clustering results for the volume limited samples. Thus we assume that the results do not suffer from incompleteness in stellar mass due to magnitude limiting.

\citeauthor{Li_2006} find increasing clustering amplitude as a function of stellar mass, with a sharp increase above the characteristic stellar mass
($\sim10^{10.6}~h^{-2}\,\mathrm{M}_\odot$), which is similar to the clustering trend they find when selecting galaxies by their $r$-band luminosity.

\subsection{Galaxy and Mass Assembly Survey}
\label{sec:obsdata_GAMA}

The Galaxy and Mass Assembly Survey (GAMA) is a multi-wavelength survey (far UV to far IR), with complete spectroscopy to $r=19.8$ (\citealp{Driver_2011}; Liske et al.\ in preparation; see Table \ref{tab:obspapers}). There are two clustering studies in GAMA focusing on stellar mass (Palamara et al.\ in preparation; Farrow et al.\ in preparation). For the present analysis, we compare to the clustering of galaxies in differential stellar mass bins as computed by Farrow et al.\ using the GAMA-II data. This is selected from the seventh SDSS data release \citep{Abazajian_2009}, with sky coverage of 180 square degrees, out to $z\lesssim0.5$, and spectroscopic completeness of $>98$ percent.

To estimate stellar masses, Farrow et al.\ use an empirical relation based on the observed $g-i$ colour and absolute magnitude in the $i$-band, as fitted by \cite{Taylor_2011} to galaxies from the second GAMA data release, with $r<19.4$ (Liske et al.\ in preparation).\footnote{
This release is based on GAMA-I \citep{Baldry_2010}.
}
To derive this relation, \citeauthor{Taylor_2011} implement broad-band SED fitting using the \cite{Bruzual_Charlot_2003} SPS model, with a \cite{Chabrier_2003} IMF,\footnote{
The \cite{Chabrier_2003} IMF has $x=1.3$ for $M_\star>1~\mathrm{M}_\odot$, and adopts a smooth transition below $1~\mathrm{M}_\odot$ to a slope similar to the \cite{Kroupa_2001} IMF (see Eqn.\ \ref{eq:imf}).
}
and exponentially declining SFHs. Extinction by dust is modelled using the \cite{Calzetti_2000} law. The photometric bands used in the fitting are the GAMA $u$, $g$, $r$, $i$, and $z$.  \citeauthor{Taylor_2011} use likelihood-weighting of all template SEDs, which can suppress discreteness effects due to the lack of interpolation between the small number of metallicities typically available in SPS models, improving on the common practice of taking the mode of the likelihood distribution (e.g.\ \citealp{Mitchell_2013}).

Farrow et al.\  use samples selected in redshift and stellar mass to compute the correlation function. We compare to their intermediate redshift range (see Table \ref{tab:obspapers}), considering the stellar mass range $10^{9.5}<M_\star~[h^{-2}\,\mathrm{M}_\odot]<10^{11.5}$. They find that clustering amplitude increases with stellar mass. A decrease in clustering strength with redshift is noted for masses below $10^{10.5}~h^{-2}\,\mathrm{M}_\odot$, with no significant redshift evolution above this.

\subsection{VIMOS Public Extragalactic Redshift Survey}
\label{sec:obsdata_VIPERS}

The VIMOS Public Extragalactic Redshift Survey (VIPERS) consists of spectroscopic observations of galaxies selected using Canada-France-Hawaii Telescope Legacy Survey (CFHTLS) photometry \citep{Guzzo_2014}. \cite{Marulli_2013} present galaxy clustering as a function of stellar mass in the first VIPERS data release \citep{Garilli_2014}. This dataset has sky coverage of about 11 square degrees, with spectroscopic completeness of roughly 40 percent. The redshift range sampled is $0.5<z<1.2$, and the magnitude limit is as given in Table \ref{tab:obspapers}. A selection in colour is used to exclude galaxies with $z<0.5$, which is not completely efficient at selecting galaxies towards the lower limit of the surveyed redshift range, i.e.\ the selection does not correspond exactly to a step function at $z=0.5$ \citep{Garilli_2014}. However, in practice this has little impact on the number of galaxies recovered across the full redshift range, and is not important for our comparisons, according to tests with the model galaxies (see \S\ref{sec:analysis}).

\citeauthor{Marulli_2013} use broad-band SED fitting to estimate the stellar masses of their sample of VIPERS galaxies, as described by \cite{Davidzon_2013}. The SPS model of \cite{Bruzual_Charlot_2003} is used, with a \cite{Chabrier_2003} IMF; i.e.\ the same choices as made by \cite{Taylor_2011} for GAMA (see \S\ref{sec:obsdata_GAMA} above). Both the \cite{Calzetti_2000} and Prevot-Bouchet \citep{Prevot_1984,Bouchet_1985} dust attenuation laws are used,\footnote{
The Prevot-Bouchet law results from modelling the dust attenuation of the Small Magellanic Cloud, while the Calzetti law was calibrated using a sample of starburst galaxies.
}
with the best-fitting option being chosen for each galaxy. Both exponentially declining and constant SFHs are used in the fitting. The photometric bands used are: the CFHTLS $u$, $g$, $r$, $i$, and $z$; the Wide-Field Infra-Red Camera (WIRCAM; \citealp{Puget_2004}) $K$; the Galaxy Evolution Explorer (GALEX; \citealp{Paz_2007}) far-UV and near-UV; the UKIRT Infra-Red Telescope Deep Sky Survey (UKIDSS; \citealp{Lawrence_2007}) $Y$, $J$, $H$, and $K$; and the Spitzer Wide-Area Infra-Red Extragalactic Survey (SWIRE; \citealp{Lonsdale_2003}) 3.6 and 4.5 $\mu$m.

Samples selected in redshift and stellar mass are used by \citeauthor{Marulli_2013} to compute the correlation function. The clustering strength increases with stellar mass in each redshift range. We compare the model clustering predictions to the lower redshift interval considered by \citeauthor{Marulli_2013} (see Table \ref{tab:obspapers}), for stellar masses $M_\star>10^{9.5}~h^{-2}\,\mathrm{M}_\odot$.

\section{Analysis}
\label{sec:analysis}

Our analysis consists of two distinct components: (i) computation of the real space clustering as a function of stellar mass in the Gon14 and Lac14 \textsc{galform} models at redshifts of 0.089, 0.32, and 0.62, and (ii) comparison of the projected model clustering as a function of stellar mass to observational data at redshifts of 0.089, 0.17, and 0.62 (i.e.\ the MS-W7 snapshots closest to the median redshifts of the observed galaxy samples; see Table \ref{tab:obspapers}). In (i), the lower and upper redshifts are the same as those in (ii), for comparing to SDSS and VIPERS data respectively; however the intermediate redshift (0.32) is chosen to be roughly evenly spaced in lookback time between these redshifts, covering 1.2 to 5.9 Gyr, rather than using the redshift of the comparison to GAMA (0.17). In (ii), we carry out SED fitting to obtain stellar mass estimates for the model galaxies, to allow a more robust comparison to the observations. The predictions obtained using the new subhalo dynamical friction merger scheme are considered in both parts of our analysis.

For each model galaxy, \textsc{galform} outputs the true stellar mass, real space coordinates, peculiar velocity, and photometry including dust attenuation (see \S \ref{sec:gf_overview}). Redshift space coordinates are computed by taking the line of sight as the third Cartesian axis and taking into account the peculiar motions along this axis (this is the distant observer approximation for projected clustering).

We convert the stellar masses predicted by \textsc{galform} from units of $h^{-1}\,\mathrm{M}_\odot$ to $h^{-2}\,\mathrm{M}_\odot$ using $h=0.704$ (see \S\ref{sec:gf_simulation}), to be consistent with the mass units of the observational data.

Apparent magnitude limits were imposed to match those of each survey, at certain stages in our analysis which will be indicated in \S\ref{sec:analysis_sm_schfun} and \S\ref{sec:analysis_clustering_pcf}. These made use of SDSS $r$-band filter wavelength response data to match GAMA, and CFHTLS $i$-band filter response data to match VIPERS (see Table \ref{tab:obspapers}). In matching SDSS, we did not impose (apparent or absolute; cf.\ Table \ref{tab:obspapers}) magnitude limits, as the clustering results of \cite{Li_2006} correspond to galaxy samples which are volume limited (see \S\ref{sec:obsdata_SDSS}).

VIPERS uses a selection in colour to exclude galaxies with $z<0.5$, as noted in \S\ref{sec:obsdata_VIPERS}, which leads to incomplete sampling of galaxies close to $z=0.5$. \cite{Guzzo_2014} note that the erroneous exclusion of galaxies ceases for $z\gtrsim0.6$. We have verified that this selection is indeed unimportant for the models by $z=0.62$.

This section describes our adjustments to the model galaxy stellar masses (\S\ref{sec:analysis_sm}), and the computation of the correlation functions used to describe their clustering (\S\ref{sec:analysis_clustering}).

\subsection{Stellar masses}
\label{sec:analysis_sm}

When computing clustering predictions to compare to observations, we considered three sets of model stellar masses: (i) the true stellar masses as predicted by \textsc{galform}, (ii) estimates of the masses from SED fitting to the model photometry, and (iii) masses resulting from abundance matching to the stellar mass functions reported for the observed galaxies (as an adjustment following the SED fitting). In particular, we present the clustering results of the models with the standard merger scheme using (i),  (ii), and (iii), and the results of the new subhalo dynamical friction merger scheme using (iii). The determination of (ii) and (iii) will now be described.

\subsubsection{SED fitting}
\label{sec:analysis_sm_sed}

\cite{Mitchell_2013} implemented SED fitting (see \S\ref{sec:obsdata_SED}) to estimate stellar masses for model galaxies, using broad-band photometry. These stellar mass estimates, when compared to the true values calculated in the model, can be used to investigate the influences of the various assumptions which are required in SED fitting (e.g.\ differences in the choice of IMF, the SFH, recycling of stellar mass back into the interstellar medium, the metallicities available in the SPS models and dust attenuation), on the derived properties of the galactic population.

Following \citeauthor{Mitchell_2013}, we carried out SED fitting for both \textsc{galform} models, at each redshift, in order to obtain estimates of the stellar masses of the model galaxies. As noted in \S \ref{sec:obsdata}, the GAMA and VIPERS stellar masses are themselves derived through broad-band SED fitting (the GAMA stellar masses use an empirical formula based on this; \citealp{Taylor_2011}), so our intention here is to carry out equivalent fitting procedures for the model galaxies. The SDSS stellar masses were derived by fitting to particular spectral features, rather than to broad-band photometry (see \S\ref{sec:obsdata_SDSS}); we nonetheless carried out broad-band SED fitting for this comparison.

In all cases the \cite{Bruzual_Charlot_2003} SPS model was assumed, with a \cite{Chabrier_2003} IMF, exponentially declining SFHs, and the \cite{Calzetti_2000} dust extinction law. The metallicities used to compute the template SEDs were matched to those used in the GAMA and VIPERS SED fitting by \cite{Taylor_2011} and \cite{Davidzon_2013} respectively, at the relevant redshifts. \citeauthor{Taylor_2011} use the full native metallicity grid of the \citeauthor{Bruzual_Charlot_2003} SPS model, i.e.\ $Z\in\lbrace 0.0001, 0.0004, 0.004, 0.008, 0.02, 0.05 \rbrace$, while \citeauthor{Davidzon_2013} use $Z\in\lbrace 0.004, 0.02 \rbrace$. To compare to SDSS, the full set of metallicity values was employed in the fitting, as in the comparison to GAMA.  Each template SED had a non-evolving metallicity, and we did not interpolate between the available SPS model metallicities, in keeping with the typical observational SED fitting procedures. All the photometric bands listed in \S \ref{sec:obsdata_GAMA} and \S \ref{sec:obsdata_VIPERS} were used in the SED fitting to match GAMA and VIPERS respectively, with the exception of the UKIDSS $K$-band, which was only included by \citeauthor{Davidzon_2013} in the absence of WIRCAM $K$-band magnitudes for VIPERS. The standard SDSS $u$, $g$, $r$, $i$, and $z$ filter set was used in the fitting to match SDSS. \citeauthor{Taylor_2011} compute likelihood-weighted stellar masses (see \S\ref{sec:obsdata_GAMA}). We conformed to this for the fitting to match GAMA, but used the more standard approach of selecting the SED at the mode of the likelihood distribution for the fitting to match SDSS and VIPERS.

\subsubsection{Schechter function matching}
\label{sec:analysis_sm_schfun}

The logarithmic stellar mass function of galaxies $\Phi(M_\star)$ is defined such that the number of galaxies per unit volume with stellar mass in the range $\log_{10}(M_\star/M_\mathrm{ref})$ to $\log_{10}(M_\star/M_\mathrm{ref}) + \mathrm{d}\log_{10}(M_\star/M_\mathrm{ref})$ is $\mathrm{d}n=\Phi(M_\star) \,\mathrm{d}\log_{10}(M_\star/M_\mathrm{ref})$, for some mass unit $M_\mathrm{ref}$. This is conventionally described using a \cite{Schechter_1976} function, where $\tilde{M}_\star$ is a characteristic mass, $\alpha$ is a power law slope, and $\tilde{\Phi}$ is a normalisation,
\begin{equation}
\mathrm{d}n
= \ln(10) \tilde{\Phi} \left( \frac{M_\star}{\tilde{M}_\star} \right) ^{\alpha + 1}
\exp \left( - \frac{M_\star}{\tilde{M}_\star} \right)
\,\mathrm{d}\log_{10}\left(\frac{M_\star}{M_\mathrm{ref}}\right)
~.
\label{eq:schfn}
\end{equation}

Differences between the numbers of model and observed galaxies in a given mass interval may give rise to discrepancies in the clustering results for samples selected by stellar mass, even if the underlying clustering signal is identical. Such mass function differences are dependent on the details of the model physics, combined with the procedure for estimating stellar masses through SED fitting.  In order to eliminate any differences between the model and observationally inferred stellar mass functions, the \citeauthor{Schechter_1976} functions representing the stellar mass functions of the galaxies in each observational sample (see Table \ref{tab:obspapers}) were imposed on the model galaxies at the corresponding redshifts. This process is equivalent to rescaling the estimated stellar mass of each \textsc{galform} galaxy, in order to make the model mass functions match the observational results. The procedure used was to match the shape and normalisation of the target \citeauthor{Schechter_1976} function, while maintaining the ordering of the model galaxies in estimated stellar mass from SED fitting, as follows. Given the lowest stellar mass of interest (for the clustering samples, see \S\ref{sec:obsdata}), and the \citeauthor{Schechter_1976} function fitted to the measurement for the observed galaxies, we generated the expected number of galaxies in the simulation volume by randomly sampling stellar masses consistent with this mass function. The generated masses were assigned in order to the \textsc{galform} galaxies, such that the highest generated mass was ascribed to the \textsc{galform} galaxy with the highest mass estimate from SED fitting, and so on. It is important to note that the galaxy formation models considered here have not been calibrated to reproduce observationally inferred stellar mass function data. They have however, been calibrated to match the local $K$-band luminosity function.

Measuring $\Phi(M_\star)$ requires one to know the number of galaxies within some stellar mass range of interest, in some known volume. Difficulties in achieving this arise from the fact that galaxy surveys are defined by apparent magnitude limits; that is, a sample of survey galaxies defined by stellar mass limits will inherently also be restricted to some range in apparent magnitude. It is, however, still possible to identify samples of galaxies which are complete, i.e.\ volume limited, in stellar mass (e.g.\ the clustering samples of \citealp{Li_2006}). The level of incompleteness in stellar mass, i.e.\ the fraction of missing galaxies, in a sample defined by stellar mass, depends on the apparent magnitude limits of the survey and the redshift range of interest.

In the typical case, where we are concerned with a faint apparent magnitude limit, it is useful to be able to estimate a lower mass threshold above which the measured mass function can be considered to be `reliable' (i.e.\ the same as what would have been measured with no magnitude limit, for a volume limited sample). \cite{Pozzetti_2010} describe the method used by \cite{Davidzon_2013} to estimate such a threshold mass for VIPERS. At our comparison redshift (see Table \ref{tab:obspapers}), this VIPERS threshold value is approximately $10^{9.6}~h^{-2}\,\mathrm{M}_\odot$, which corresponds to the mass below which the faint limit causes more than about 20 percent of galaxies to be missed, in a given mass interval. The VIPERS \citeauthor{Schechter_1976} function specified in Table \ref{tab:obspapers} has been fitted by \citeauthor{Davidzon_2013} only for masses above the reliability threshold defined in this way.  Clearly the $i$-band faint limit has an important influence on the completeness of the stellar mass function at this redshift, for the lowest masses of interest (note that the minimum mass we consider for the clustering calculations using these galaxies is $10^{9.5}~h^{-2}\,\mathrm{M}_\odot$). In light of this, we imposed the $i$-band faint limit on the model galaxies \textit{before} matching the VIPERS mass function at this redshift, in order to reproduce the `underestimated' mass function measurement at lower masses, i.e.\ the incomplete mass function of the galaxies actually used in the clustering analysis of \cite{Marulli_2013}.

The double \citeauthor{Schechter_1976} function representing the GAMA stellar mass function has been fitted for $z<0.06$ \citep{Baldry_2012}, whereas the clustering results of Farrow et al.\ (in preparation) considered here use galaxies with $0.14< z < 0.24$ (see Table \ref{tab:obspapers}). \citeauthor{Baldry_2012} use the SED fitting procedure of \cite{Taylor_2011} to estimate stellar masses, supplemented by the corresponding empirical relation of \citeauthor{Taylor_2011} for a small number of galaxies with missing photometry (see \S\ref{sec:obsdata_GAMA}). \citeauthor{Baldry_2012} compare their stellar mass function to that obtained by \cite{Pozzetti_2010} using the Cosmic Evolution Survey (zCOSMOS; \citealp{Lilly_2007}) for $0.1<z<0.35$, using similar SED fitting for the stellar mass estimation, and find good agreement between the two measurements. Furthermore, we have verified that the stellar mass function does not evolve significantly when using the sample of galaxies considered by Farrow et al.\ ($0.14< z < 0.24$), with respect to the \citeauthor{Baldry_2012} measurement. The \citeauthor{Baldry_2012} mass function fit can be regarded as complete for the masses of interest here, thanks to being constrained at low redshift (the fitting considered the mass function as measured for $M_\star>10^{7.7}~h^{-2}\,\mathrm{M}_\odot$, while the clustering data we compare to are for $M_\star>10^{9.5}~h^{-2}\,\mathrm{M}_\odot$). Thus, we matched the mass functions of the \textsc{galform} models to the double \citeauthor{Schechter_1976} fit, while keeping all the model galaxies, and only \textit{afterwards} imposed the $r$-band faint limit on the models (see Table \ref{tab:obspapers}). In this way, the complete model stellar mass functions were made to reproduce the complete observationally inferred mass function, before introducing the relative incompleteness due to the particular selection relevant for comparison to the clustering data of Farrow et al.

The SDSS mass function \citeauthor{Schechter_1976} fit was derived directly from the second SDSS data release \citep{Abazajian_2004,Wang_2006}, using $r$-band apparent magnitude and redshift limits the same as for the clustering data of \cite{Li_2006}, but without the additional absolute magnitude limits imposed for the clustering samples (see Table \ref{tab:obspapers}). We assumed the \citeauthor{Wang_2006} fit to be approximately complete in stellar mass, and thus imposed it upon the model galaxies without applying any magnitude limiting, as required for comparison to the volume limited clustering samples of \citeauthor{Li_2006} (see \S\ref{sec:obsdata_SDSS}). It should be noted, however, that the measured mass function is likely to be significantly suppressed by the SDSS faint $r$-band limit for $M_\star\lesssim10^{8.2}~h^{-2}\,\mathrm{M}_\odot$ \citep{Baldry_2008,Li_White_2009}. This incompleteness has minor implications for our comparison to the \citeauthor{Wang_2006} mass function data and \citeauthor{Schechter_1976} fit (for $M_\star>10^{8.69}~h^{-2}\,\mathrm{M}_\odot$) in \S\ref{sec:results_comp}, but is not important for our comparison to the volume limited clustering data; except where the mass function is imposed, in which case any errors introduced by the assumption of completeness in the \citeauthor{Schechter_1976} fit should only have an impact on the lowest masses considered.

In \S\ref{sec:results_comp_matching} and \S\ref{sec:results_comp_sdf} we make use of the mass function matching to compare the results of the two \textsc{galform} merger schemes to the observational clustering data (as outlined at the beginning of \S\ref{sec:analysis_sm}). In this way, we force the models using either scheme to reproduce the same mass function, and thus the differences in the clustering results using the two schemes, when compared in this way, are not due to any differences in the predicted stellar mass function which are introduced by changing to the new merger scheme.

\subsection{Clustering}
\label{sec:analysis_clustering}

We now define the correlation function $\xi(r)$, and the projected correlation function $w_\mathrm{p}(\sigma)$, and give details of their computation.

\subsubsection{Correlation function}
\label{sec:analysis_clustering_cf}

For a cosmologically representative volume $V$, the probability $\mathrm{d}P$ of finding a galaxy in a volume element $\mathrm{d}V_1$, at a comoving distance $r_{12}$ from another galaxy in a volume element $\mathrm{d}V_2$, defines the spatial two-point autocorrelation function $\xi(r)$ such that,
\begin{equation}
\mathrm{d}P = n_V^2 [1 + \xi(r_{12})] \,\mathrm{d}V_1 \,\mathrm{d}V_2 ~,
\label{eq:xi_def}
\end{equation}
where $n_V$ is the mean number density of galaxies within $V$ \citep{Peebles_1980}.
Thus following  \cite{Rivolo_1986},
\begin{equation}
1 + \xi (r) = \frac{1}{n_V^2 V} \frac{N(r)}{V_\mathrm{s}(r)} ~,
\label{eq:xi_calc}
\end{equation}
where $N(r)$ is the number of pairs with separation $r$, computed by summing over the number of pairs involving each galaxy in the volume individually (so this is twice the number of independent pairs), considering a spherical shell of radius $r$ and volume $V_\mathrm{s}(r)$. A simple appreciation of the uncertainty in $\xi(r)$ can be gained from considering Poisson statistics \citep{Iovino_1988, Martinez_1993}. The error on $\xi(r)$ is then $\Delta\xi(r)$ such that,
\begin{equation}
\Delta\xi(r) = \left(\frac{2}{n_V^2 V } \frac{1 + \xi(r)}{V_\mathrm{s}(r)} \right) ^\frac{1}{2}~.
\label{eq:xi_err}
\end{equation}

It is common to fit a power law to $\xi(r)$, parametrized by a correlation length $r_0$ and slope $\gamma$ \citep{Peebles_1980},
\begin{equation}
\xi(r) = \left(\frac{r}{r_0}\right)^{- \gamma} ~.
\label{eq:xi_fit}
\end{equation}
Here $r_0$ characterises the clustering `strength', where $\xi(r_0)=1$. This parameterization is not suitable for describing the precise clustering measurements which are possible today over a wide range of scales, but we can make use of it to describe the galaxy clustering over small ranges of pair separations.

$\xi(r)$ was computed for the models at redshifts $z\in\lbrace0.089, 0.32, 0.62\rbrace$ (using the true stellar masses, with no magnitude limits), with and without the new merger scheme, in three equally spaced bins in stellar mass. Pair separations in the range $0.1 < r~[h^{-1}\,\mathrm{Mpc}] < 30$ (the choice of the large-scale limit will be discussed in \S\ref{sec:analysis_clustering_limit}) were divided into 30 bins of equal logarithmic width. $\xi(r)$ was computed using the pair counts binned in $r$, following Eqn.\ (\ref{eq:xi_calc}), where $V=L_\mathrm{box}^3$ (see \S\ref{sec:gf_simulation}).

The power law given in Eqn.\ (\ref{eq:xi_fit}) was fitted to each $\xi(r)$, using the $\Delta\xi(r)$ from Eqn.\ (\ref{eq:xi_err}) to weight the fit. We preferred to fit over a relatively small range in $r$, where $\xi(r)$ is close to being an exact power law, in the neighbourhood of $\xi(r)=1$, such that $r_0$ relates closely to the `true' correlation length. In this way, bins within $3 < r~[h^{-1}\,\mathrm{Mpc}] < 10$ were used to fit the power law for each $\xi(r)$. This range in $r$ samples the two-halo term in the correlation function, i.e.\ the separations considered relate to galaxies in different haloes. As such, the fitted power laws are insensitive to clustering on small scales (one-halo term), and thus we consider them using only the standard merger scheme.

Our results for $\xi(r)$ using the two \textsc{galform} models, comparing the two merger schemes, are presented in \S\ref{sec:results_models}.

\begin{figure}
\centering
\includegraphics[width=240pt]{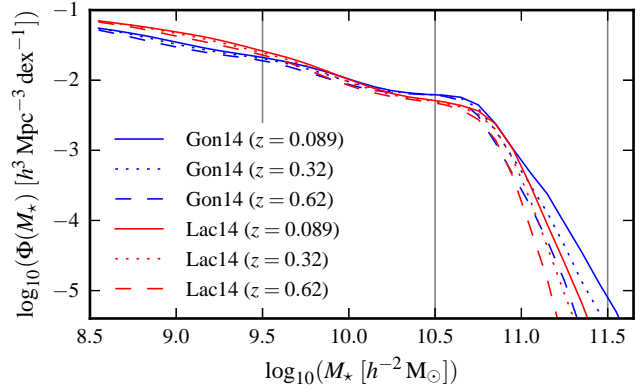}
\caption{
Galactic stellar mass function $\Phi(M_\star)$ predicted by the Gon14 and Lac14 \textsc{galform} models, as a function of redshift. The vertical lines indicate the stellar mass samples used to measure $\xi(r)$ in Fig.\ \ref{fig:xi}. The lower $\Phi(M_\star)$ axis limit corresponds to 50 model galaxies per bin in the MS-W7 simulation volume.
}
\label{fig:smf}
\end{figure}

\begin{figure*}
\centering
\includegraphics[width=500pt]{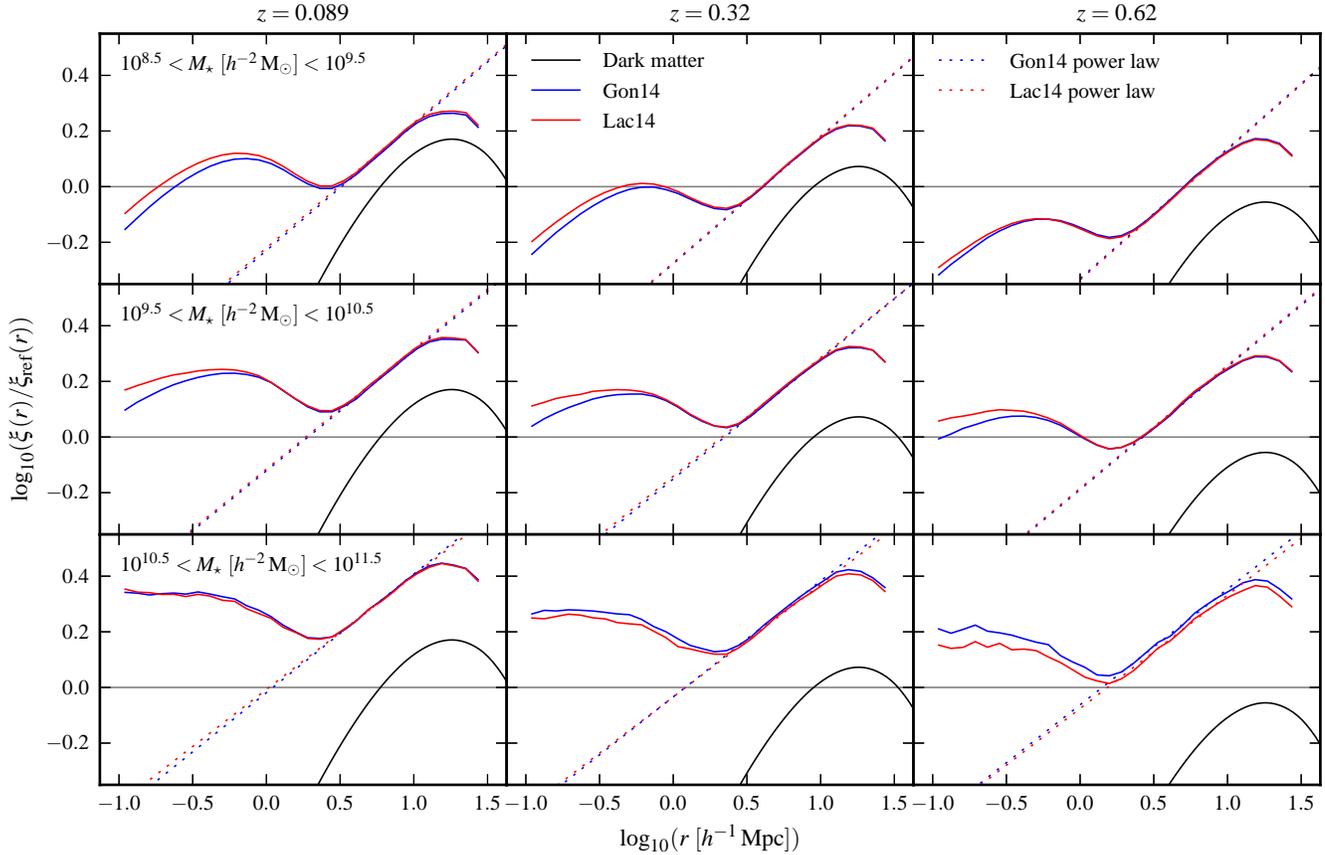}
\caption{
Predicted correlation function of galaxies $\xi(r)$ as a function of redshift $z$ (columns) and stellar mass $M_\star$ (rows), computed using the Gon14 and Lac14 \textsc{galform} models. For clarity, each $\xi(r)$ has been divided by a reference power law $\xi_\mathrm{ref}(r)$, with parameters $r_0 = 5~h^{-1}\,\mathrm{Mpc}$ and $\gamma = 2$ (horizontal line; see Eqn.\ \ref{eq:xi_fit}). The dotted lines are power law fits to each \textsc{galform} $\xi(r)$, for $3 < r~[h^{-1}\,\mathrm{Mpc}] < 10$, with the parameters shown in Fig.\ \ref{fig:powerlaws}. The black curves show $\xi(r)$ for the dark matter at each redshift, computed as the Fourier transform of the MS-W7 linear theory power spectrum.
}
\label{fig:xi}
\end{figure*}

\begin{figure}
\centering
\includegraphics[width=240pt]{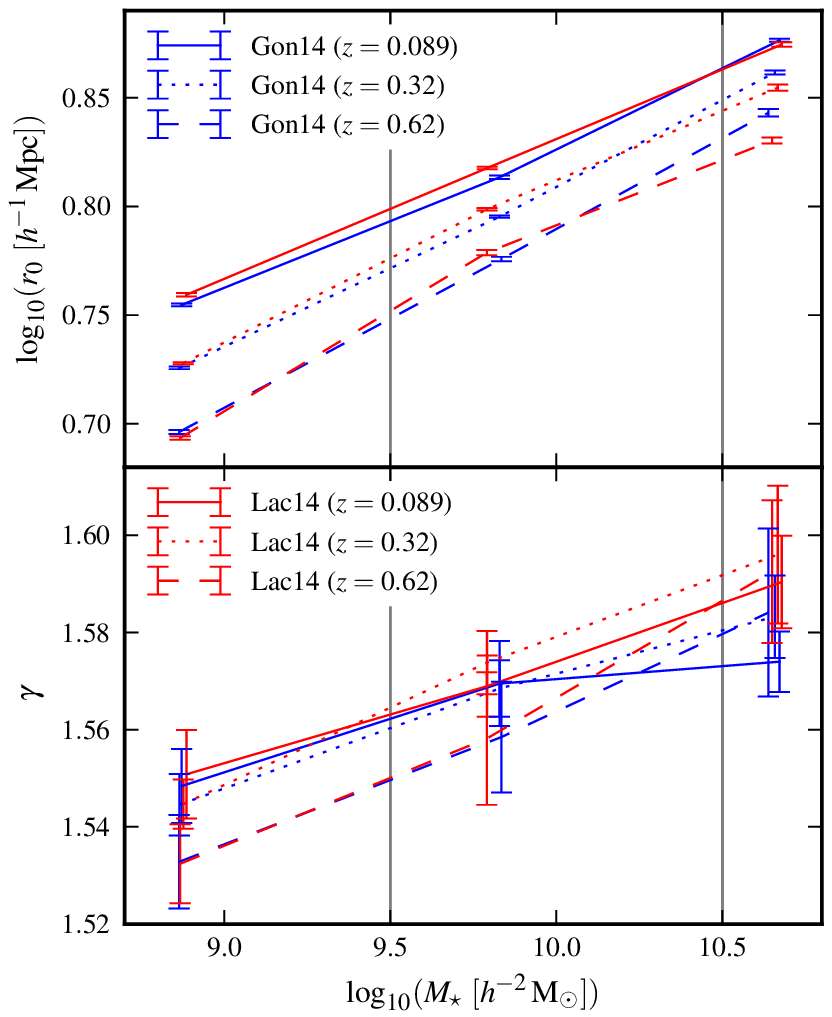}
\caption{
Correlation lengths $r_0$ (upper) and slopes $\gamma$ (lower) of the power law fitted to the $\xi(r)$ shown in Fig.\ \ref{fig:xi} for the Gon14 and Lac14 \textsc{galform} models, as a function of stellar mass $M_\star$ and redshift $z$ (see Eqn.\ \ref{eq:xi_fit}). The vertical lines indicate the divisions between the mass bins used to measure $\xi(r)$. The $M_\star$ values are the medians for each mass bin, model, and redshift. Each panel uses the same linestyles and colours; the legend is split between panels.
}
\label{fig:powerlaws}
\end{figure}

\subsubsection{Projected correlation function}
\label{sec:analysis_clustering_pcf}

The distance to a galaxy moving with the cosmic expansion can be inferred through measuring its recession velocity (redshift) and using Hubble's law. If the galaxy has some peculiar velocity relative to the Hubble flow, this will perturb the determined distance. As such, it is necessary to distinguish between real space and redshift space coordinates, i.e.\ true spatial positions, and those determined through measuring recession velocities, respectively \citep{Kaiser_1987}.

For observed galaxies, the redshift space correlation function may be computed. This encodes information about the peculiar motions of galaxies on different separation scales. Through considering the components of the pair separation orthogonal and parallel to the line of sight ($\sigma$ and $\pi$ respectively), we can define the two-dimensional correlation function $\xi(\sigma, \pi)$,\footnote{
We reserve $\xi(r)$ to mean the spherically averaged two-point correlation function computed using real space pair separations $r$. We use ($\sigma$, $\pi$) to refer to the pair separation components in \textit{either} real or redshift space, and state where $w_\mathrm{p}(\sigma)$ has been computed from real or redshift space coordinates (i.e.\ real or redshift space values of the line of sight separation $\pi$).
}
by analogy with Eqn.\ (\ref{eq:xi_calc}).
We may integrate $\xi(\sigma, \pi)$ along the line of sight, defining the projected correlation function $w_\mathrm{p}(\sigma)$ \citep{Davis_Peebles_1983},
\begin{equation}
w_\mathrm{p}(\sigma) = \int_{-\pi_\mathrm{max}}^{\pi_\mathrm{max}} \xi(\sigma, \pi) \, \mathrm{d}\pi ~.
\label{eq:wp_def}
\end{equation}
This statistic is traditionally used to describe the real space clustering of galaxies drawn from observational surveys, as redshift space distortions do not influence $w_\mathrm{p}(\sigma)$. This is true provided that $\pi_\mathrm{max}$ is sufficiently large, such that the integral includes all correlated galaxy pairs, encompassing their redshift-distorted coordinates \citep{Davis_Peebles_1983, Norberg_2009}. \citeauthor{Norberg_2009} show using simulations that $w_\mathrm{p}(\sigma)$ is sensitive to the choice of $\pi_\mathrm{max}$, with differences between the projected clustering recovered on large scales for $\pi_\mathrm{max}~[h^{-1}\,\mathrm{Mpc}]\in\lbrace30,64\rbrace$. They note that using $\pi_\mathrm{max}= 64~h^{-1}\,\mathrm{Mpc}$ (cf.\ Table \ref{tab:obspapers}) results in a difference of ten percent between real and redshift space for $\sigma\sim10~h^{-1}\,\mathrm{Mpc}$, rising to 50 percent by $\sigma\sim30~h^{-1}\,\mathrm{Mpc}$.

Assuming Eqn.\ (\ref{eq:xi_fit}) for $\xi(r)$, and infinite $\pi_\mathrm{max}$, the corresponding power law for $w_\mathrm{p}(\sigma)$ is given by \citep{Davis_Peebles_1983},
\begin{equation}
w_\mathrm{p}(\sigma) = \sigma^{1-\gamma} r_0^\gamma
\frac{\Gamma(1/2)\Gamma([\gamma - 1]/2)}{\Gamma(\gamma/2)} ~.
\label{eq:wp_fit}
\end{equation}
This result is used in this paper only for illustrative purposes.

$w_\mathrm{p}(\sigma)$ was computed in real and redshift space for the two \textsc{galform} models at each redshift $z\in\lbrace0.089, 0.17, 0.62\rbrace$, for each of the four variants identified in \S\ref{sec:analysis_sm}, i.e.\ the default models with their true and estimated masses, and the models with masses resulting from matching the observational mass functions (using both the new and default merger schemes). At each redshift the appropriate magnitude limits were imposed (see the beginning of \S\ref{sec:analysis}). The stellar mass binning was matched to that of each survey. Separations orthogonal to the line of sight were computed with the same binning as the $r$ values for $\xi(r)$ (see \S\ref{sec:analysis_clustering_cf}). Line of sight separations were measured in bins of width $\Delta \pi = 1~h^{-1}\,\mathrm{Mpc}$, using the distant observer approximation (as noted at the beginning of \S\ref{sec:analysis}). $\xi(\sigma, \pi)$ was evaluated for each bin of the $\sigma$ versus $\pi$ grid, using Eqn.\ (\ref{eq:xi_calc}), where the volume $V_\mathrm{s}$ considered was instead that of a cylindrical shell of radii set by the $\sigma$ bin edges and length $\Delta \pi$. Discretizing Eqn.\ (\ref{eq:wp_def}), $w_\mathrm{p}(\sigma) = \sum_{\pi_i} \xi(\sigma,\pi_i)\, \Delta\pi$ was computed for bin centres $\pi_i$, such that $\left| \pi_i \right| < \pi_\mathrm{max}$ ($30~h^{-1}\,\mathrm{Mpc}$, see \S\ref{sec:analysis_clustering_limit} below).

The results of our comparison of the model clustering to the observational $w_\mathrm{p}(\sigma)$ data are presented in \S\ref{sec:results_comp}.

\subsubsection{The limit on separation scales}
\label{sec:analysis_clustering_limit}

The finite size of the simulation volume sets an upper limit on the separation scales where the galaxy clustering can be considered to be reliable, with respect to that which would be computed using an arbitrarily large cosmological volume.

To establish the pair separations where the box size becomes important, we computed the projected correlation function along three mutually orthogonal lines of sight. This was carried out for each redshift and stellar mass range of interest in this paper, in both real and redshift space.

We found the scatter in $w_\mathrm{p}(\sigma)$ for $\sigma<30~h^{-1}\,\mathrm{Mpc}$, around the mean for the three projections, to be about 0.01 dex. For $\sigma\gtrsim30~h^{-1}\,\mathrm{Mpc}$, the scatter rose sharply with increasing $\sigma$, reaching roughly 0.1 dex by $\sigma\sim50~h^{-1}\,\mathrm{Mpc}$. These results for the $w_\mathrm{p}(\sigma)$ scatter across projections are largely insensitive to the choice of line of sight integration limit, over the range $30 < \pi_\mathrm{max} ~[h^{-1}\,\mathrm{Mpc}]< 50$.

Based on this test, we used an upper limit of $30~h^{-1}\,\mathrm{Mpc}$ for $\sigma$ when computing $w_\mathrm{p}(\sigma)$, and for $r$ when computing $\xi(r)$. $\pi_\mathrm{max}=30~h^{-1}\,\mathrm{Mpc}$ was used when comparing to each survey. We have checked that using the particular $\pi_\mathrm{max}$ values given in Table \ref{tab:obspapers} for each survey instead (which are larger for the SDSS and GAMA measurements) does not have a significant impact on our comparisons. Our upper limit to pair separations is the same as that noted by \cite{Orsi_2008} as the largest separation where the galactic and dark matter $\xi(r)$ are related by a scale independent bias in the Millennium Simulation volume. \cite{Gonzalez_2011} also found this scale to be the upper limit for describing the redshift space correlation function boost with the \cite{Kaiser_1987} formalism, due to the finite size of the simulation box.

\begin{figure}
\centering
\includegraphics[width=240pt]{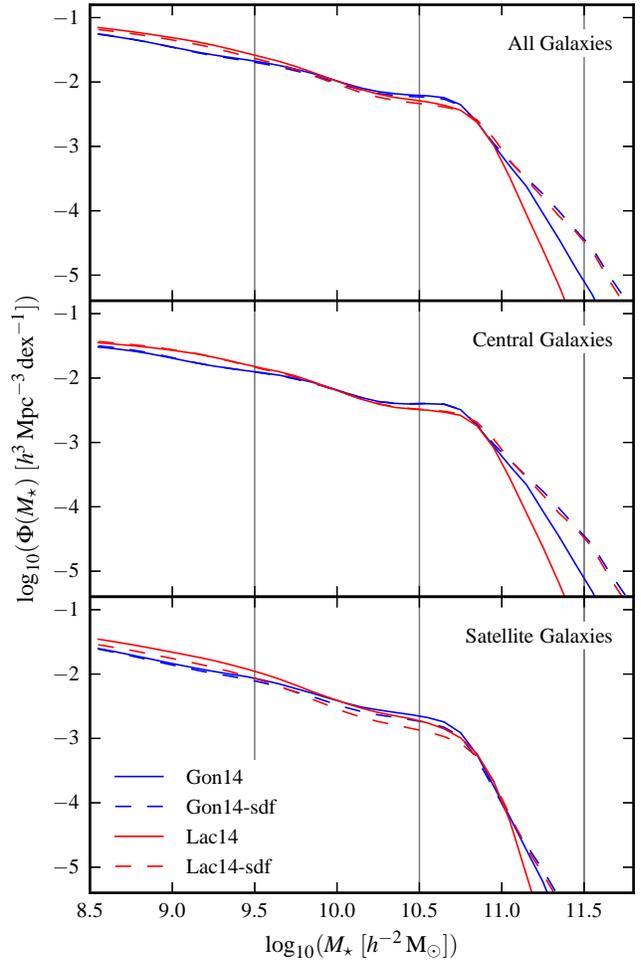}
\caption{
Galactic stellar mass function $\Phi(M_\star)$ at redshift $z=0.089$, for all (top), central (middle), and satellite (bottom) galaxies in the Gon14 and Lac14 \textsc{galform} models, and for these models with the alternative subhalo dynamical friction merger scheme (labelled Gon14-sdf and Lac14-sdf). The vertical lines indicate the stellar mass samples used to compute $\xi(r)$ in Fig.\ \ref{fig:xi_sdf}. The lower $\Phi(M_\star)$ axis limit corresponds to 50 model galaxies per bin in the MS-W7 simulation volume. Note that the model parameters have not been recalibrated on adopting the new merger scheme.
}
\label{fig:smf_sdf}
\end{figure}

\begin{figure*}
\centering
\includegraphics[width=500pt]{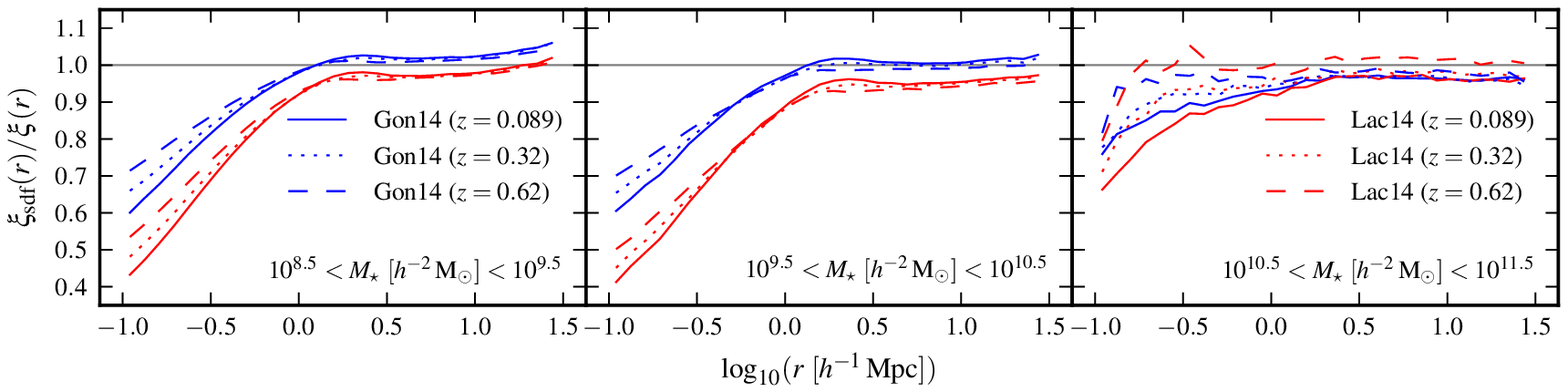}
\caption{
Ratio of the real space correlation function using the new subhalo dynamical friction merger scheme, $\xi_\mathrm{sdf}(r)$, to that obtained with the default satellite merger prescription, $\xi(r)$, computed using the Gon14 and Lac14 \textsc{galform} models, as a function of stellar mass and redshift (as labelled). The linestyles show different redshifts and the colours indicate different models as labelled, with the legend split across panels.
}
\label{fig:xi_sdf}
\end{figure*}

\section{Results and Discussion}
\label{sec:results}

Our results are now presented for the real space $\xi(r)$ computed using the \textsc{galform} models as a function of true stellar mass and redshift (\S\ref{sec:results_models}), and the comparisons to the $w_\mathrm{p}(\sigma)$ observational clustering estimates (\S\ref{sec:results_comp}).

\subsection{Predicted real space clustering}
\label{sec:results_models}

We first consider the results of the default models (\S\ref{sec:results_models_std}), and then examine the influence of the choice of satellite merger scheme on the small-scale clustering (\S\ref{sec:results_models_sdf}).

\subsubsection{Standard models}
\label{sec:results_models_std}

The galactic stellar mass function $\Phi(M_\star)$ as a function of redshift is shown in Fig.\ \ref{fig:smf}, as predicted using the Gon14 and Lac14 \textsc{galform} models. Fig.\ \ref{fig:xi} shows the real space $\xi(r)$ computed as a function of stellar mass and redshift. Fig.\ \ref{fig:powerlaws} shows the parameters $r_0$ and $\gamma$ of the power laws fitted to each $\xi(r)$ bin.

As shown in Fig.\ \ref{fig:smf}, the evolution of the shape of $\Phi(M_\star)$ as a function of redshift is fairly similar between the models. The number density above the knee of the mass function decreases with increasing redshift in each model. There are more high-mass galaxies in the Gon14 model than in the Lac14 model (this is similar to the differences between the predicted $K$-band luminosity functions). This regime roughly corresponds to the highest mass bin for the correlation functions shown in Fig.\ \ref{fig:xi} (as indicated in Fig.\ \ref{fig:smf}).

For clarity, we divide the predicted correlation functions by a reference power law (see Fig.\ \ref{fig:xi} caption). The model clustering predictions shown in Fig.\ \ref{fig:xi} are close to a power law over only a limited range of scales. There is a clear transition between the one-halo term on small scales and the two-halo term on large scales (at $r\sim2~h^{-1}\,\mathrm{Mpc}$). The two-halo term has the same shape as the large-scale dark matter $\xi(r)$, where the galaxy clustering bias with respect to the dark matter increases with both redshift and stellar mass. The clustering in the two models is indistinguishable on large scales (i.e.\ for pair counts between haloes), but there are differences between the model predictions on smaller scales ($r\lesssim1~h^{-1}\,\mathrm{Mpc}$).

There is a general trend of clustering amplitude increasing with stellar mass, on all separation scales, at each redshift. Slight increases in the amplitude of the clustering on large scales are seen with decreasing redshift, for a fixed mass bin. The power law fits shown in Fig.\ \ref{fig:xi} ignore clustering on small and very large scales. The fit parameters, as shown in Fig.\ \ref{fig:powerlaws}, indicate increasing clustering strength $r_0$ with both increasing stellar mass and decreasing redshift, where the differences between redshifts become less significant for higher stellar masses. The slope $\gamma$ is fairly constant, but exhibits a weak increase with increasing stellar mass. The redshift evolution of the clustering on small scales in Fig.\ \ref{fig:xi} (which is sensitive to pair separations within haloes) is particularly dramatic for low stellar masses.

\subsubsection{New satellite merger scheme}
\label{sec:results_models_sdf}

 In the new merger scheme, satellite galaxies track the positions of their associated subhaloes, only making use of an analytically computed merger time-scale once the subhalo can no longer be resolved in the simulation, as described in \S\ref{sec:gf_sdf}. Fig.\ \ref{fig:smf_sdf} shows the stellar mass functions of the Gon14 and Lac14 models at redshift $z=0.089$, with and without the new merger scheme, decomposed into the contributions from central and satellite galaxies. The same trends are seen at higher redshifts. Changes in the total $\Phi(M_\star)$ due to the new merger scheme are significant for the highest masses (top panel of Fig.\ \ref{fig:smf_sdf}), where for both models the amplitude of the high-mass end increases when switching to the new scheme. Note that the model parameters have not been recalibrated on adopting the new merger scheme. The rise in the galaxy abundance in the Lac14 model at these masses is greater than for the Gon14 model, such that with the new scheme, the two model mass functions become very similar for high masses. The increases in the numbers of the most massive galaxies are due to more mergers taking place in the new scheme, which thus transfer more mass to central galaxies, as can be seen in the middle panel of Fig.\ \ref{fig:smf_sdf}. There is a smaller rise at high masses in the satellite mass function for both models, which again is more significant for the Lac14 model (lower panel). The source of the growth of higher mass galaxies here is the earlier assimilation of lower mass satellites. Mergers on to a galaxy cease when it becomes a satellite of a more massive galaxy.

The changes in $\xi(r)$ for the two \textsc{galform} models when using the new merger scheme are displayed in Fig.\ \ref{fig:xi_sdf}, at redshifts of 0.089, 0.32, and 0.62. The lines show the results with the subhalo dynamical friction merger scheme, divided by the predictions using the original merger scheme, for each model, stellar mass interval, and redshift. The clustering differences with respect to the standard merger scheme are larger for the lower redshifts. Generally there is little change in the clustering on large scales. There are significant decreases in the small-scale clustering amplitude ($r\lesssim1~h^{-1}\,\mathrm{Mpc}$) for all but the highest mass, highest redshift, data. The intermediate mass bin shows the largest decreases in $\xi(r)$ on small scales, although these are similar to the reductions in amplitude for the lower mass bin. There is significantly less change in the highest mass bin, relative to that in the other mass bins, at a given redshift. Changes in both the radial distribution and number of satellites contribute to the changes in the small-scale clustering when using the new scheme (see lower panel of Fig.\ \ref{fig:smf_sdf}, and \S\ref{sec:gf_sdf}).

\begin{figure}
\centering
\includegraphics[width=240pt]{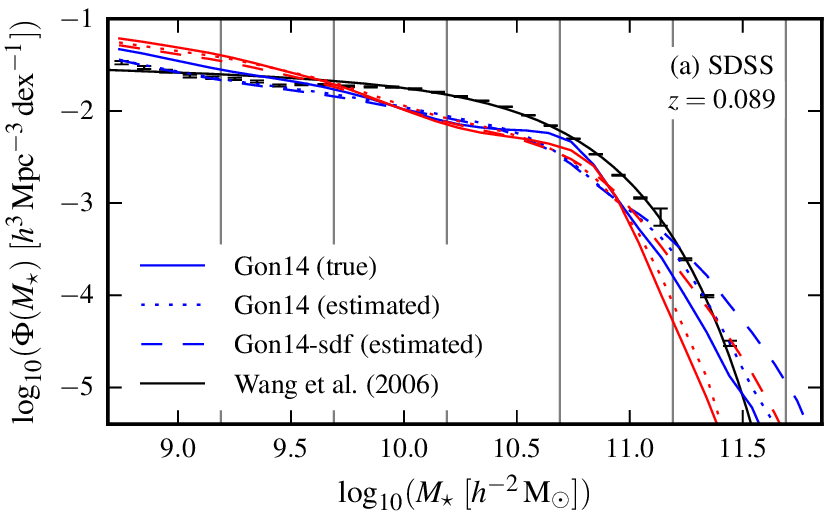}
\includegraphics[width=240pt]{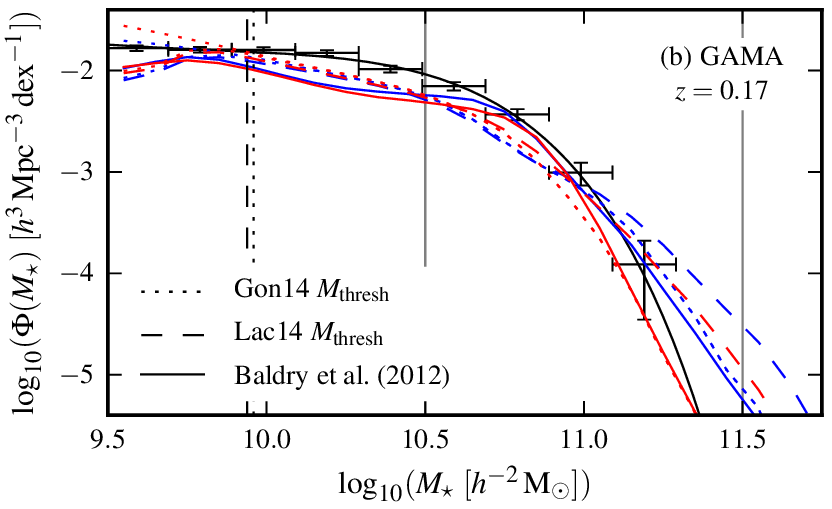}
\includegraphics[width=240pt]{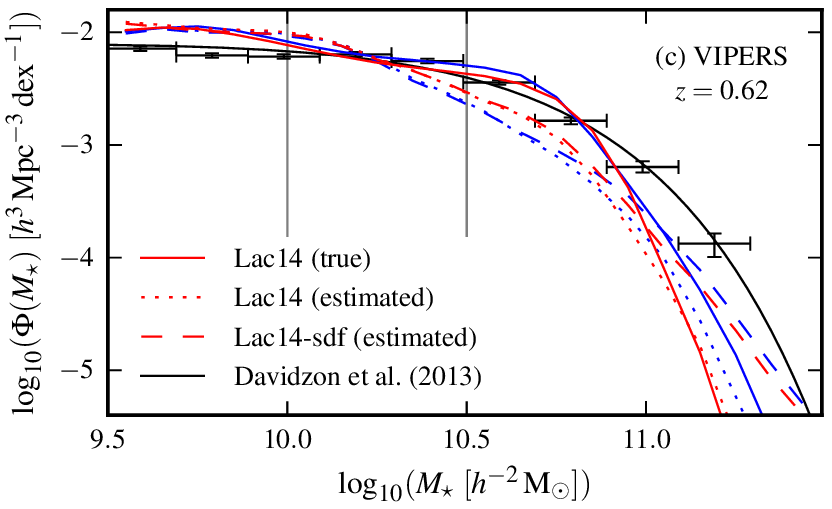}
\caption{
Galactic stellar mass function $\Phi(M_\star)$, as predicted using the Gon14 and Lac14 \textsc{galform} models, at redshifts (a) 0.089, (b) 0.17, and (c) 0.62.
Solid and dotted lines show the true and estimated $\Phi(M_\star)$ respectively for the default models. Dashed lines show the estimated $\Phi(M_\star)$ for the models using the new merger scheme (Gon14-sdf and Lac14-sdf). The formats of the model lines are the same in each plot, and are labelled in (a) and (c).
The points with error bars are the measurements of (a) \protect\citet{Wang_2006} using SDSS, (b) \protect\citet{Baldry_2012} using GAMA, and (c) \protect\citet{Davidzon_2013} using VIPERS; shown alongside \protect\citet{Schechter_1976} function fits (black lines; see Table \ref{tab:obspapers}).
In (b) and (c), the apparent magnitude limits of Farrow et al.\ (in preparation) and \protect\citet{Marulli_2013} have been imposed on the model galaxies, respectively (see Table \ref{tab:obspapers} and \S\ref{sec:analysis}).
The solid vertical grey lines in (a), (b), and (c) indicate the samples used to measure $w_\mathrm{p}(\sigma)$ in Figs.\ \ref{fig:wp_SDSS}, \ref{fig:wp_GAMA}, and \ref{fig:wp_VIPERS} respectively. The lower $\Phi(M_\star)$ axis limit corresponds to 50 model galaxies per bin in the MS-W7 simulation volume.
At $z=0.17$, we compare to the $z<0.06$ \protect\citeauthor{Baldry_2012} measurement under the assumption that there is little evolution in $\Phi(M_\star)$ since $z=0.17$ (see \S\ref{sec:analysis_sm_schfun}).
The \protect\citeauthor{Baldry_2012} $\Phi(M_\star)$ measurement is complete over the mass range shown in (b), and should not be compared to the magnitude limited model $\Phi(M_\star)$ at the lowest masses. The estimated $\Phi(M_\star)$ for the default models is repeated in (b) without magnitude limiting (with the same formatting), to show the impact of the faint limit at this redshift.
To aid comparison, a corresponding threshold mass $M_\mathrm{thresh}$ is indicated for each default model in (b) by a dotted or dashed vertical black line (see \S\ref{sec:results_comp_true_est}), as labelled in (b).
}
\label{fig:smf_obs}
\end{figure}

\subsection{Comparisons to observational data}
\label{sec:results_comp}

We now present the results of our comparisons to observational measurements of the clustering of galaxy samples selected by stellar mass. First we consider the changes induced in the model predictions through stellar mass estimation using SED fitting (\S\ref{sec:results_comp_true_est}), then discuss rescaling of the estimated model mass functions to reproduce observational mass function measurements (\S\ref{sec:results_comp_matching}), and present the changes in the clustering predictions resulting from using the new subhalo dynamical friction merger scheme in \textsc{galform} (\S\ref{sec:results_comp_sdf}). In \S\ref{sec:results_comp_needSED} we comment on the need for SED fitting in order to make a fair comparison to observational clustering measurements, even when carrying out mass function abundance matching. Finally, we discuss some uncertainties relevant to these comparisons to observational clustering results (\S\ref{sec:results_comp_uncert}).

\subsubsection{True and estimated stellar masses}
\label{sec:results_comp_true_est}

\cite{Mitchell_2013} show that for some ranges in stellar mass, the process of SED fitting to the broad-band photometry of model galaxies, as predicted by \textsc{galform}, introduces a roughly constant offset in estimated stellar mass with respect to true model masses, combined with associated scatter (see \S\ref{sec:analysis_sm_sed}). However, for \textsc{galform} galaxies with strong dust attenuation, SED fitting tends to significantly underestimate the stellar mass, leading to important systematic differences between the distributions of true and estimated masses, which cannot be described by a constant mean offset and scatter. This phenomenon is relevant for model galaxies with true stellar mass $\gtrsim10^{10}~h^{-2}\,\mathrm{M}_\odot$.

It is important to note that the IMF assumed in a galaxy formation model influences mass-to-light ratios. The predicted luminosity functions are directly sensitive to the IMF, and are used to calibrate the models, which in turn affects the stellar masses. Within the machinery of SED fitting, the choice of IMF corresponds to a roughly constant offset in the resulting stellar masses.

The \cite{Chabrier_2003} IMF used in our SED fitting matches the GAMA and VIPERS stellar masses, but not the \cite{Kroupa_2001} IMF of the SDSS data (see \S \ref{sec:obsdata} and \S \ref{sec:analysis_sm_sed}). Therefore we multiplied the mass estimates by 1.1 following the SED fitting to match SDSS. This factor is that found by \cite{Davidzon_2013} as the mean offset between the stellar masses derived using \citeauthor{Chabrier_2003} and \citeauthor{Kroupa_2001} IMFs in their SED fitting (see also \citealp{Baldry_2008}).

The stellar mass functions of the Gon14 and Lac14 models are shown in Fig.\ \ref{fig:smf_obs}, at redshifts of 0.089, 0.17, and 0.62, alongside the relevant observationally inferred mass function data from SDSS, GAMA, and VIPERS, respectively. The true and estimated mass functions are shown for each model using solid and dotted lines, respectively (the dashed lines show the estimated mass functions using the new merger scheme, which will be discussed in \S\ref{sec:results_comp_sdf}). For the comparison to SDSS (Fig.\ \ref{fig:smf_obs}a), no magnitude limits have been imposed on the model galaxies, because the clustering measurements of \cite{Li_2006} are for samples which are complete in stellar mass (see \S\ref{sec:obsdata_SDSS}). When comparing to GAMA and VIPERS (Fig.\ \ref{fig:smf_obs}b,c), we imposed the apparent magnitude limits relevant for comparison to the clustering data of Farrow et al.\ (in preparation) and \cite{Marulli_2013} respectively (see Table \ref{tab:obspapers} and \S\ref{sec:analysis}).

The \cite{Baldry_2012} GAMA stellar mass function fit can be regarded as complete in stellar mass over the range of masses shown in Fig.\ \ref{fig:smf_obs}b, as it was constrained for $z<0.06$ (see \S\ref{sec:analysis_sm_schfun}). Therefore we also show the estimated model stellar mass functions \textit{without} applying the GAMA magnitude limit in Fig.\ \ref{fig:smf_obs}b, as these are appropriate to compare to the \citeauthor{Baldry_2012} data over the full range of masses shown (these lines are shown with the same formatting as the magnitude limited versions). Thus the differences due to the magnitude limit can be seen where the dotted lines of a given colour do not overlap (the lower, magnitude limited, dotted lines are close to the solid true model stellar mass lines at the lowest masses, following a reduction in amplitude of roughly 0.4 dex due to the magnitude limit). In this way, it can be seen that the faint $r$-band limit of GAMA influences the completeness of the estimated model stellar mass functions for $M_\star\lesssim10^{10}~h^{-2}\,\mathrm{M}_\odot$ at $z=0.17$. The vertical (dotted or dashed) black lines in Fig.\ \ref{fig:smf_obs}b indicate a threshold stellar mass $M_\mathrm{thresh}$ for each model, above which the GAMA magnitude limit causes one percent of the galaxies in the complete model sample to be missed in the simulation volume, computed using the estimated stellar masses.

\begin{figure*}
\centering
\includegraphics[width=500pt]{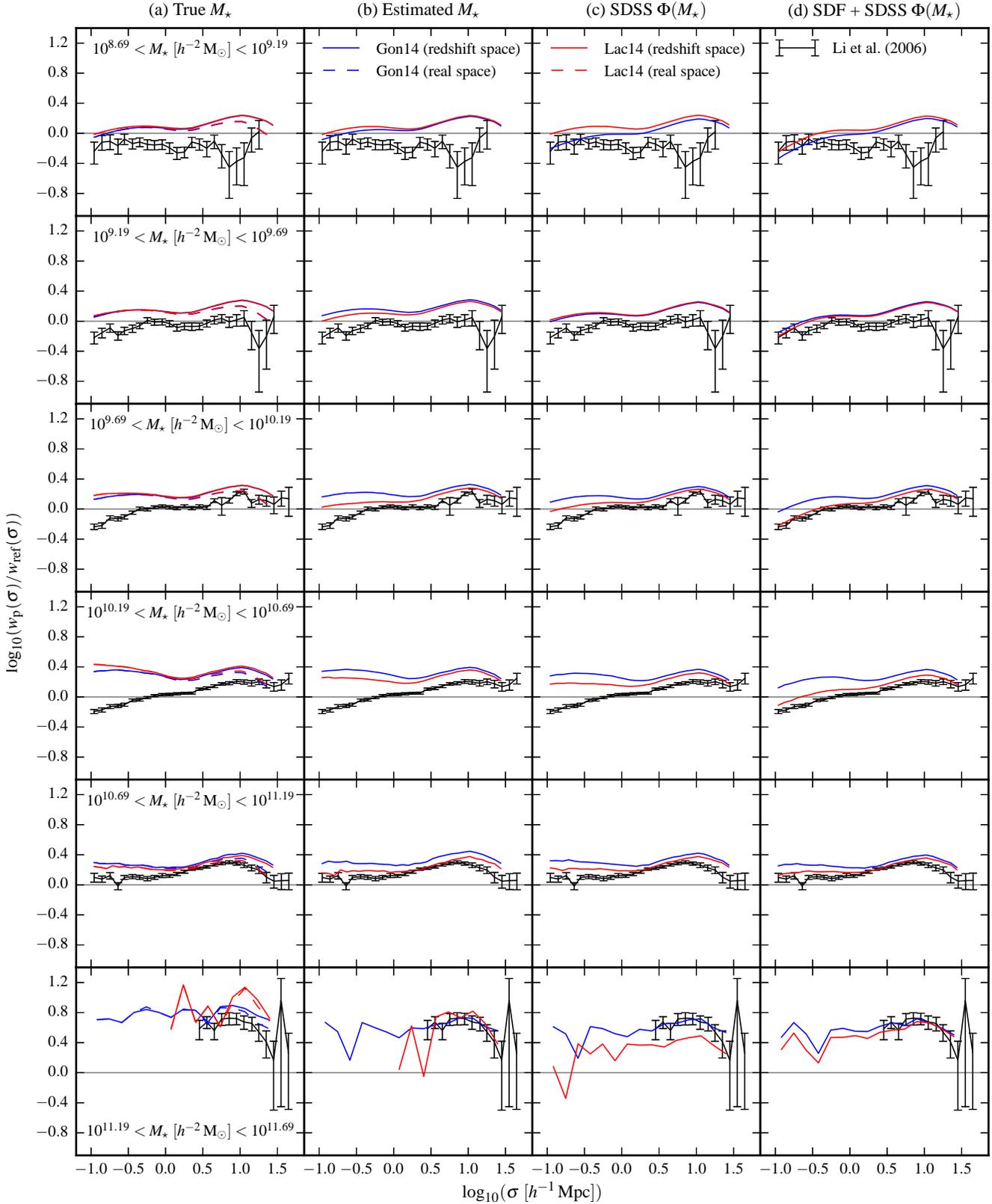}
\caption{
Projected correlation function of galaxies $w_\mathrm{p}(\sigma)$ as a function of stellar mass $M_\star$ (as labelled in each row) at redshift $z=0.089$, computed using the Gon14 and Lac14 \textsc{galform} models, in real (first column only) and redshift space. The SDSS measurements of \protect\citet{Li_2006} are shown (black lines with errorbars). Each $w_\mathrm{p}(\sigma)$ has been divided by a reference power law $w_\mathrm{ref}(\sigma)$, with parameters $r_0 = 5~h^{-1}\,\mathrm{Mpc}$ and $\gamma = 2$ (horizontal line; see Eqn.\ \ref{eq:wp_fit}). Half the standard number of $\sigma$ bins are shown for the highest mass interval, due to the small number of galaxies. Columns (a) and (b) show the default models with the true and estimated masses, respectively. The remaining columns show the results of matching the SDSS stellar mass function (as an adjustment following the SED fitting), for (c) the default models, and (d) the models using the new subhalo dynamical friction (SDF) merger scheme.
}
\label{fig:wp_SDSS}
\end{figure*}

\begin{figure*}
\centering
\includegraphics[width=500pt]{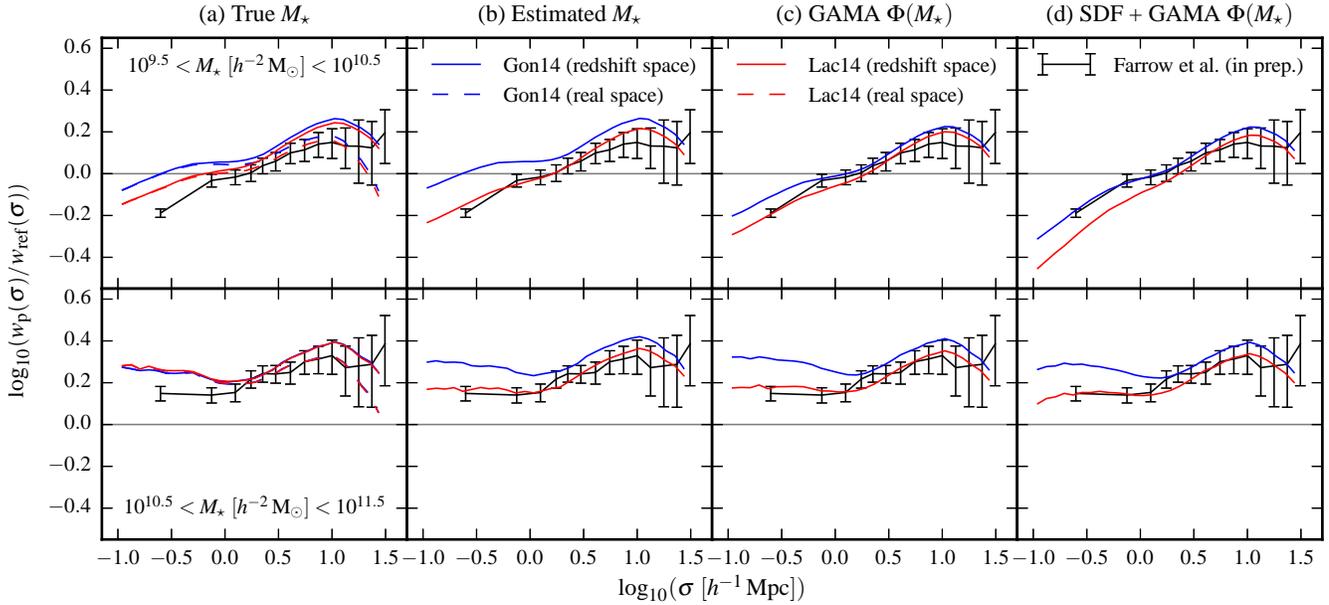}
\caption{
Projected correlation function of galaxies $w_\mathrm{p}(\sigma)$ for different stellar mass $M_\star$ bins (as labelled in each row) at redshift $z=0.17$, computed using the Gon14 and Lac14 \textsc{galform} models, in real (first column only) and redshift space. The GAMA measurements of Farrow et al.\ (in preparation) are shown (black lines with errorbars), whose apparent magnitude limit has been imposed on the model galaxies. For clarity, each $w_\mathrm{p}(\sigma)$ has been divided by a reference power law $w_\mathrm{ref}(\sigma)$, with parameters $r_0 = 5~h^{-1}\,\mathrm{Mpc}$ and $\gamma = 2$ (horizontal line; see Eqn.\ \ref{eq:wp_fit}). Columns (a) and (b) show the default models with the true and estimated masses, respectively. The remaining columns show the results of matching the GAMA stellar mass function (as an adjustment following the SED fitting), for (c) the default models, and (d) the models using the new subhalo dynamical friction (SDF) merger scheme.
}
\label{fig:wp_GAMA}
\end{figure*}

\begin{figure*}
\centering
\includegraphics[width=500pt]{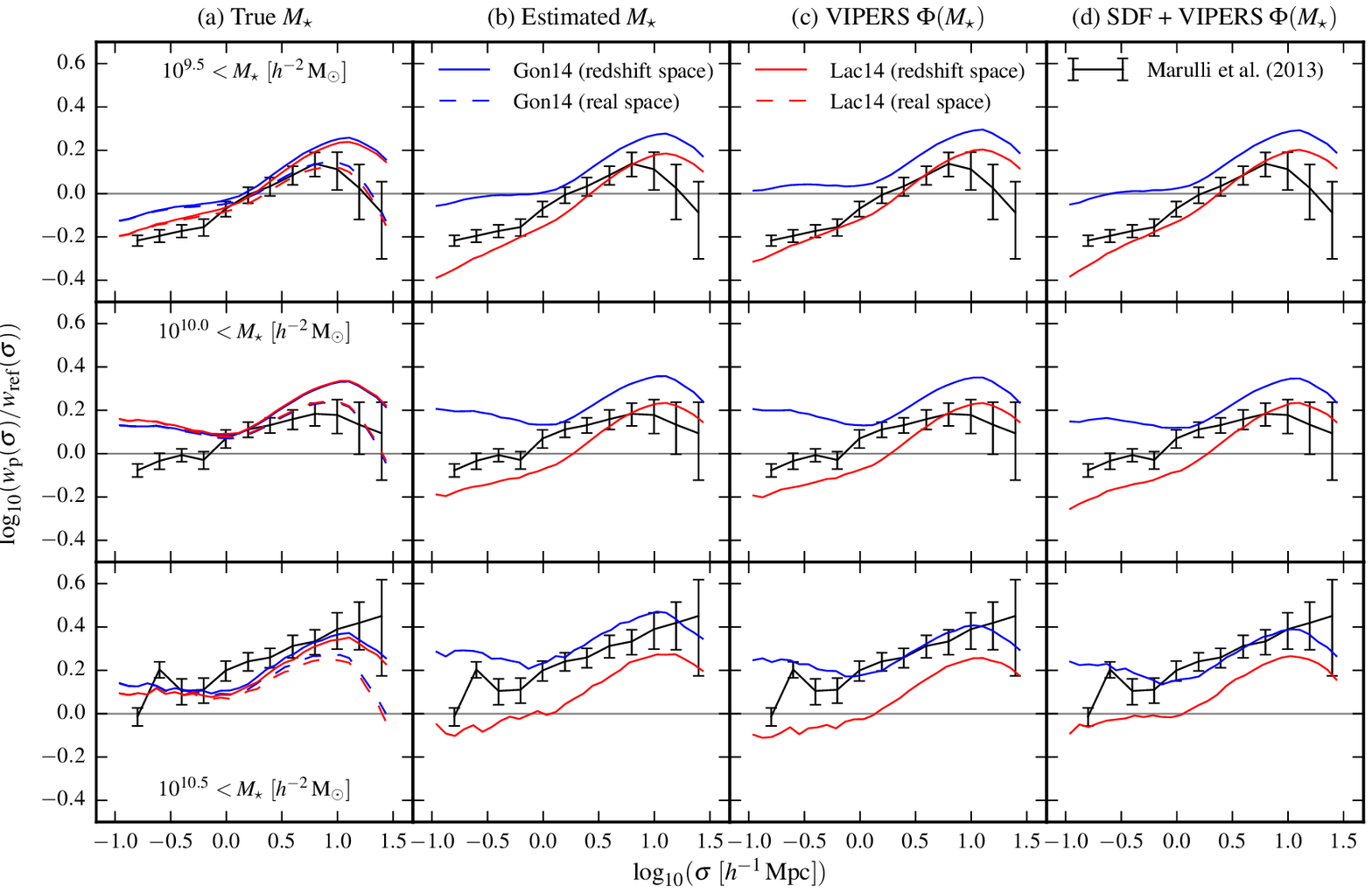}
\caption{
Projected correlation function of galaxies $w_\mathrm{p}(\sigma)$ as a function of cumulative stellar mass $M_\star$ (as labelled in each row) at redshift $z=0.62$, computed using the Gon14 and Lac14 \textsc{galform} models, in real (first column only) and redshift space. The VIPERS measurements of \protect\citet{Marulli_2013} are shown (black lines with errorbars), whose apparent magnitude limit has been imposed on the model galaxies. For clarity, each $w_\mathrm{p}(\sigma)$ has been divided by a reference power law $w_\mathrm{ref}(\sigma)$, with parameters $r_0 = 5~h^{-1}\,\mathrm{Mpc}$ and $\gamma = 2$ (horizontal line; see Eqn.\ \ref{eq:wp_fit}). Columns (a) and (b) show the default models with the true and estimated masses, respectively. The remaining columns show the results of matching the VIPERS stellar mass function (as an adjustment following the SED fitting), for (c) the default models, and (d) the models using the new subhalo dynamical friction (SDF) merger scheme.
}
\label{fig:wp_VIPERS}
\end{figure*}

The true model stellar mass functions can change significantly when carrying out SED fitting. An effect which can be seen at each redshift is that the knee of the estimated mass function is smoothed out and suppressed with respect to the true mass function, for both models. The changes at the knee become larger with increasing redshift, and can be traced to the influence of dust attenuation (see \citealp{Mitchell_2013}). Additionally, scatter in the estimated masses at a given true mass will cause $\Phi(M_\star)$ to rise for masses above this, particularly near the high-mass end, due to the rapid changes in number density as a function of stellar mass. This Eddington bias conspires with the systematic effects of dust attenuation to dictate the shape of the estimated mass function.

The projected correlation functions computed using the model galaxies, for samples selected by stellar mass, are shown in Figs.\ \ref{fig:wp_SDSS}, \ref{fig:wp_GAMA}, and \ref{fig:wp_VIPERS}, alongside the relevant observational measurements, for the comparisons to SDSS, GAMA, and VIPERS, respectively. In each case, column (a) shows the model clustering predictions when using the true stellar masses, and column (b) shows the results when using the estimated masses from SED fitting. These masses correspond to the true and estimated mass functions shown in Fig.\ \ref{fig:smf_obs}, where the relevant $w_\mathrm{p}(\sigma)$ mass bin edges are indicated by vertical lines. The results shown in columns (c) and (d) of Figs.\ \ref{fig:wp_SDSS}, \ref{fig:wp_GAMA}, and \ref{fig:wp_VIPERS} will be discussed in \S\ref{sec:results_comp_matching} and \S\ref{sec:results_comp_sdf} respectively.

Switching from true to estimated model stellar masses can change the clustering predictions significantly, comparing columns (a) and (b) in Figs.\ \ref{fig:wp_SDSS}, \ref{fig:wp_GAMA}, and \ref{fig:wp_VIPERS}. More massive haloes are more strongly clustered in hierarchical cosmologies (e.g.\ \citealp{Cole_1989}), and the stellar mass of a galaxy is closely related to its host halo mass (e.g.\ \citealp{Guo_2010}). Therefore the changes in the clustering predictions when carrying out SED fitting can be understood in terms of model galaxies in dark matter haloes of certain masses being transferred across the stellar mass bin boundaries. The Gon14 and Lac14 models predict relatively similar clustering in most cases when using true stellar masses, but the SED fitting tends to significantly increase the differences between the results for the two models. The changes in the clustering predictions of the Lac14 model when switching to estimated masses are often larger than for the Gon14 model. It is likely that this is due to a combination of different factors, such as the different SPS models assumed and levels of dust extinction calculated in the two models. The SED fitting tends to decrease the clustering amplitude in the Lac14 model, but tends to increase that predicted in the Gon14 model. In general, switching to estimated masses has a larger impact on the clustering on small scales ($\sigma\lesssim1~h^{-1}\,\mathrm{Mpc}$), such that the shape of $w_\mathrm{p}(\sigma)$ is modified.

Comparing to SDSS, the agreement with the clustering measurements of \cite{Li_2006} is improved or similar when switching from true to estimated stellar masses, for both \textsc{galform} models (see Fig.\ \ref{fig:wp_SDSS}a,b). For both models in most mass bins, the clustering amplitudes using true or estimated masses are higher over the range of separations shown than inferred from SDSS. However, there is somewhat better agreement for the highest masses, in particular, when using estimated masses in the second highest (for Lac14) and highest (for Gon14 and Lac14) mass bins.

In Fig.\ \ref{fig:wp_GAMA}a, the clustering predictions using true stellar masses are similar for the two models, except that the Lac14 model predicts relatively low clustering amplitude in the lower mass bin. There is reasonable agreement with the large-scale GAMA clustering measurements of Farrow et al.\ (in preparation), especially in the higher mass bin. Both models predict stronger clustering on the smallest scales than is inferred observationally; this difference is most significant for the Gon14 model in the lower mass bin. The Lac14 model clustering predictions yield improved agreement with the GAMA measurements when switching to estimated masses. However, the corresponding increases in the clustering amplitudes of the Gon14 model lead to poorer agreement with the GAMA data than found with the true stellar masses (see Fig.\ \ref{fig:wp_GAMA}a,b).

Using true stellar masses, the model predictions agree with the VIPERS clustering measurements of \cite{Marulli_2013} on intermediate scales ($\sigma\sim1~h^{-1}\,\mathrm{Mpc}$) in the two lower mass bins in Fig.\ \ref{fig:wp_VIPERS}a, where they tend to overpredict the clustering amplitude on smaller and larger scales. In the highest mass bin, both models agree reasonably well with the observational measurements on small and large scales, but predict lower clustering amplitude than observed on intermediate scales. Switching to the estimated masses (see Fig.\ \ref{fig:wp_VIPERS}b), improves the large-scale agreement with the VIPERS data for the Lac14 model, in the two lower mass bins. Otherwise, the level of agreement between the model predictions and observational measurements on large scales is similar or reduced when switching to the estimated masses. On small scales, switching to estimated masses worsens the agreement with the VIPERS clustering data for each model and mass bin, except for the Lac14 model in the intermediate bin, where the level of agreement is slightly increased. These changes result in the models with estimated masses yielding clustering predictions which encompass the VIPERS data on small scales, with the clustering amplitude being lower for the Lac14 model, and higher for the Gon14 model.

It is clear from the results shown in columns (a) and (b) of Figs.\ \ref{fig:wp_SDSS}, \ref{fig:wp_GAMA}, and \ref{fig:wp_VIPERS} that carrying out SED fitting to the broad-band photometry of model galaxies to recover estimates of their stellar masses can have a significant impact on the model clustering predictions for samples selected by stellar mass. Thus it is important to implement this procedure to obtain a reliable comparison between the clustering of galaxies as a function of stellar mass, as measured for observed galaxies and predicted by theoretical models.

\subsubsection{Abundance matching of stellar mass functions}
\label{sec:results_comp_matching}

Figs.\ \ref{fig:wp_SDSS}c, \ref{fig:wp_GAMA}c, and \ref{fig:wp_VIPERS}c show the projected correlation functions resulting from carrying out abundance matching of the estimated model stellar mass functions to \cite{Schechter_1976} function fits to the mass function data measured using each survey (SDSS, GAMA, and VIPERS respectively). The mass function fits are shown alongside the model predictions in Fig.\ \ref{fig:smf_obs}  (see \S\ref{sec:analysis_sm_schfun} for details of the matching procedure). The differences between the model clustering predictions shown in  columns (b) and (c) of Figs.\ \ref{fig:wp_SDSS}, \ref{fig:wp_GAMA}, and \ref{fig:wp_VIPERS} are thus entirely due to the model mass estimates from SED fitting being rescaled to reproduce the observationally inferred mass function fits.

Looking at the comparison to the SDSS results of \cite{Li_2006} in Fig.\ \ref{fig:wp_SDSS}b,c, the mass function matching leads to a suppression of the model clustering amplitude for both models (in most mass bins). This leads to improved, or unchanged, agreement with the observational results; except for the Lac14 model in both the second highest mass bin (where poorer agreement is found due to a rise in clustering amplitude), and in the highest mass bin (where the model $w_\mathrm{p}(\sigma)$ becomes lower in amplitude than measured from SDSS). However, the changes in the clustering predictions due to imposing the SDSS mass function are typically relatively minor, and in general there remains excessive clustering on both small and large scales with respect the SDSS measurements in Fig.\ \ref{fig:wp_SDSS}c.

Comparing to the measurements of Farrow et al.\ (in preparation) using GAMA data (see Fig.\ \ref{fig:wp_GAMA}b,c), imposing the observationally inferred mass function causes the clustering amplitude to be suppressed for both models in the lower mass bin (where the most significant impact is on the Gon14 model). This leads to improved agreement with the GAMA clustering results for the Gon14 model, and similar agreement for the Lac14 model, in this mass bin. In the higher mass bin, imposing the mass function has very little influence on the Lac14 model clustering prediction, but slightly boosts the amplitude of the small-scale clustering in the Gon14 model, exacerbating the discrepancy with the GAMA measurement on small scales.

It can be seen in Fig.\ \ref{fig:wp_VIPERS}b,c that the mass function matching leads to improved agreement with the VIPERS data in the lower mass interval for the Lac14 model, and worsened agreement for the Gon14 model, as the clustering amplitude rises in both models. The matching does not have a significant impact on the clustering predictions of either model in the intermediate mass interval. Both models exhibit suppression in the clustering amplitude due to the mass function matching for the highest masses, improving the agreement with the VIPERS data for the Gon14 model, and reducing the level of agreement for the Lac14 model. Thus there persist notable differences between the model predictions and the VIPERS measurements following the abundance matching, in Fig.\ \ref{fig:wp_VIPERS}c. The Gon14 model predicts excessive clustering with respect to the VIPERS results on both small and large scales in the two lower mass bins. The Lac14 model predicts lower clustering amplitudes than measured from VIPERS, for small scales in the intermediate mass interval, and on all scales for the highest masses.

The changes in the model clustering predictions due to matching observationally inferred stellar mass function fits reflect the underlying modifications to the distribution of halo masses in a given stellar mass interval. Decreases in the clustering amplitude are seen where there is a significant net influx of galaxies in haloes of relatively low mass, thus diluting the clustering signal (and vice versa). The changes due to the stellar mass function matching are typically smaller than the changes which occur when switching from true model stellar masses to estimated masses recovered from SED fitting to the model photometry.

\subsubsection{New satellite merger scheme}
\label{sec:results_comp_sdf}

Thus far, we have compared the model clustering predictions to observational results  using the default treatment of satellite galaxies in \textsc{galform}. In the default scheme, satellite galaxies are assumed to enter the main halo on random orbits (independently of the orbit of their associated subhalo), and merge with the central galaxy upon the elapse of an analytically determined dynamical friction merger time-scale. We now consider the results obtained when using a new merger scheme, in which satellite galaxies track their associated subhaloes until these are no longer resolved, and only then is an analytic merger time-scale computed (see \S\ref{sec:gf_sdf}).

Using the standard merger scheme (looking in particular at Figs.\ \ref{fig:wp_SDSS}c, \ref{fig:wp_GAMA}c, and \ref{fig:wp_VIPERS}c), the Gon14 model tends to predict stronger small-scale clustering than seen in the observational data. The Lac14 model predicts higher small-scale clustering than measured from SDSS in Fig.\ \ref{fig:wp_SDSS}c, but is in agreement with (or predicts lower clustering amplitude than) the small-scale results from GAMA and VIPERS (Figs.\ \ref{fig:wp_GAMA}c and \ref{fig:wp_VIPERS}c respectively). It is noteworthy that both models tend to predict higher clustering amplitudes on small scales than measured from SDSS.

Stronger small-scale clustering in the models, with respect to observational results, may be interpreted as an excess of satellite galaxies in the models, or too concentrated a radial distribution of satellites \citep{Contreras_2013}. In order to reconcile the model predictions with observations, it is possible that a more detailed study of the disruption and mergers of satellite galaxies is needed \citep{Henriques_2008,Kim_2009}. Another possibility is to make use of the substructure information available in dark matter simulations, i.e.\ to employ our new satellite merger scheme, which has an impact on small-scale clustering (see \S\ref{sec:gf_sdf} and \S\ref{sec:results_models_sdf}).

The estimated stellar mass functions of the Gon14 and Lac14 models when using the new `subhalo dynamical friction'  merger scheme are shown as dashed lines in Fig.\ \ref{fig:smf_obs} (where they are labelled Gon14-sdf and Lac14-sdf). Comparing these lines to the estimated mass functions using the standard merger scheme (dotted lines), it can be seen that the most significant differences are at the highest masses. At each redshift, both models have more high-mass galaxies when using the new scheme, where the differences between the predictions for the two schemes  for a given model emerge roughly at the knee of the mass function. For the true masses, the mass functions for both models using the new scheme are very similar to each other at the highest masses, for each survey comparison redshift (to the level shown in Fig.\ \ref{fig:smf_sdf} at the SDSS comparison redshift); and yet the estimated mass functions shown in Fig.\ \ref{fig:smf_obs} are distinct at high masses for the Gon14 and Lac14 models using the new scheme.

Figs.\ \ref{fig:wp_SDSS}d, \ref{fig:wp_GAMA}d, and \ref{fig:wp_VIPERS}d show the projected clustering predictions of the \textsc{galform} models using the new merger scheme. These results use stellar masses estimated through SED fitting, followed by matching to the observational stellar mass function fits shown in Fig.\ \ref{fig:smf_obs} (as described in \S\ref{sec:analysis_sm_schfun}). In this way, the results shown in columns (c) and (d) of Figs.\ \ref{fig:wp_SDSS}, \ref{fig:wp_GAMA}, and \ref{fig:wp_VIPERS} have been required to reproduce the same observationally inferred mass function fits, making use of the ordering of galaxies in stellar mass obtained through SED fitting, but differing in whether they use (c) the default merger scheme, or (d) the new scheme.

The results shown in Fig.\ \ref{fig:xi_sdf} show that the changes in the model clustering predictions due to the new merger scheme, for samples selected by true stellar mass, are larger for lower stellar masses and lower redshifts. This trend is also seen for the abundance matched estimated masses when comparing columns (c) and (d) in Figs.\ \ref{fig:wp_SDSS}, \ref{fig:wp_GAMA}, and \ref{fig:wp_VIPERS}.

Reductions in the clustering amplitude lead to improved small-scale agreement with the SDSS data for both models in Fig.\ \ref{fig:wp_SDSS}c,d. In the SDSS comparison, the Lac14 clustering also becomes stronger in the highest mass bin, resulting in improved agreement with the observational data on large scales. However, in several mass bins in Fig.\ \ref{fig:wp_SDSS}d the small-scale clustering amplitude is still high with respect to the SDSS measurements, particularly for the Gon14 model. Additionally, the model clustering predictions on the largest scales are often excessive with respect to the SDSS results in Fig.\ \ref{fig:wp_SDSS}d, particularly so for the two lowest mass bins.

Both models experience suppressed small-scale clustering in Fig.\ \ref{fig:wp_GAMA}c,d, which is more significant in the lower mass bin, leading to good agreement with the small-scale GAMA measurements for the Gon14 model in the lower mass bin, and for the Lac14 model in the higher mass bin. However, the Lac14 model clustering amplitude in Fig.\ \ref{fig:wp_GAMA}d is now low on small scales with respect to the GAMA result in the lower mass bin. The suppression of the Gon14 clustering in the higher mass bin, due to the new merger scheme, is small compared to the excessive clustering on small scales with respect to the GAMA measurement in Fig.\ \ref{fig:wp_GAMA}c.

In Fig.\ \ref{fig:wp_VIPERS}c,d, suppression of the small-scale clustering tends to push the $w_\mathrm{p}(\sigma)$ predicted by the Gon14 model closer to the VIPERS data, and further below this for the Lac14 model; except in the highest mass bin, where the Lac14 model clustering amplitude rises. However, the impact of the new merger scheme on the clustering predictions shown in Fig.\ \ref{fig:wp_VIPERS}d is relatively small, and the overall level of agreement with the VIPERS measurements is almost unchanged from the results shown in Fig.\ \ref{fig:wp_VIPERS}c.

\subsubsection{The need for SED fitting}
\label{sec:results_comp_needSED}

We have used abundance matching of model stellar mass functions to observational measurements as a basis for comparing the clustering predictions of the different \textsc{galform} models to each other, and to observational estimates. The abundance matching has been carried out as a final adjustment to the estimated model galaxy masses. These results are shown in columns (c) and (d) of Figs.\ \ref{fig:wp_SDSS}, \ref{fig:wp_GAMA}, and \ref{fig:wp_VIPERS}, and have been discussed in \S\ref{sec:results_comp_matching} and \S\ref{sec:results_comp_sdf} above. It is interesting to ask how these results, which have undergone mass function matching following SED fitting, compare to those obtained by simply matching the observationally inferred mass functions using the true model stellar masses. This test examines how important the SED fitting is when making clustering predictions to compare to observations, and in particular how significantly the ordering of galaxies in stellar mass changes when carrying out SED fitting (in terms of the impact on the clustering).

In a small number of cases, the changes due to the SED fitting are insignificant when using imposed mass functions. For example, the results for the lowest mass GAMA galaxies (as shown in Fig.\ \ref{fig:wp_GAMA}c,d) are unperturbed by SED fitting for both the Gon14 and Lac14 models, with or without the new merger scheme. Yet for the higher mass GAMA galaxies, not carrying out SED fitting results in clustering predictions using the new merger scheme which on small scales are midway in $\log_{10}(w_\mathrm{p}(\sigma))$ between those shown in Fig.\ \ref{fig:wp_GAMA}d, for both \textsc{galform} models. Similar trends are obtained in the highest VIPERS mass interval, for example, where for either merger scheme, both models yield small-scale clustering amplitudes midway between the model lines shown in Fig.\ \ref{fig:wp_VIPERS}c,d.

In general, we find that without SED fitting, the models with matched mass functions produce clustering predictions which are different to those of their counterparts based on estimated masses, particularly for $\sigma\lesssim1~h^{-1}\,\mathrm{Mpc}$.

\subsubsection{Sources of uncertainty}
\label{sec:results_comp_uncert}

We now discuss sources of uncertainty which are relevant to the comparisons to observational data carried out in this work.

\begin{enumerate}

\item
A potential complication when comparing to observational galaxy clustering measurements is the inability of spectrographs to resolve galaxies in close proximity on the sky (due to fibre collisions). This effect systematically lowers the clustering amplitude, particularly on small scales (e.g.\ \citealp{Zehavi_2002,Pollo_2005,delaTorre_2013}).

For SDSS, spectroscopic fibres on a given plate must be separated by at least 55 arcsec, which means that at the median redshift of \cite{Li_2006}, two galaxies cannot be observed simultaneously within $\sigma\lesssim0.1~h^{-1}\,\mathrm{Mpc}$, although this can influence $w_\mathrm{p}(\sigma)$ out to larger scales. \citeauthor{Li_2006} carry out a correction to account for this, based on comparing the projected correlation function to that derived using the parent photometric catalogue. It is possible that, despite this correction, there remains some systematic bias in the observational results on the small scales shown in Fig.\ \ref{fig:wp_SDSS}.

GAMA employs a sophisticated `greedy' tiling strategy, yielding nearly spatially uniform spectroscopic completeness \citep{Robotham_2010,Driver_2011}. Thus the small-scale clustering measurements of Farrow et al.\ (in preparation) shown in Fig.\ \ref{fig:wp_GAMA} should be reasonably reliable (i.e.\ robust against fibre collisions).

A correction is made by \cite{Marulli_2013} to account for spectroscopic fibre collisions in VIPERS, motivated by the Munich semi-analytic galaxy formation model of \cite{deLucia_Blaizot_2007}. Mock catalogues are constructed from this model, with and without modelling of the spectrograph selection. The relative difference between the correlation functions computed for the two sets of mock galaxies is used to impose a correction on the observed result. This model dependent correction may produce results which are systematically different on small scales to what would be measured for a truly spectroscopically complete sample. \cite{delaTorre_2013} note that for VIPERS the fraction of missing galaxy pairs becomes significant below 0.03 degrees, which at the median redshift of the data to which we compare corresponds to $\sigma\lesssim0.8~h^{-1}\,\mathrm{Mpc}$ (cf.\ Fig.\ \ref{fig:wp_VIPERS}).

\item
The differences on large scales between $w_\mathrm{p}(\sigma)$ computed in real and redshift space for a given model in Figs.\ \ref{fig:wp_SDSS}, \ref{fig:wp_GAMA}, and \ref{fig:wp_VIPERS} show that $\pi_\mathrm{max}=30~h^{-1}\,\mathrm{Mpc}$ is not sufficiently large for the projected correlation function computed in redshift space to converge to the real space result, but is appropriate for comparison to the observational data (see \S\ref{sec:analysis_clustering_limit}). Note that we integrate the real and redshift space pair separations out to the same $\pi_\mathrm{max}$ when computing $w_\mathrm{p}(\sigma)$ (see \S\ref{sec:analysis_clustering_pcf}). For clarity, the real space clustering is shown only in column (a) of Figs.\ \ref{fig:wp_SDSS}, \ref{fig:wp_GAMA}, and \ref{fig:wp_VIPERS}. The offsets between the real and redshift space clustering are similar for the columns where the real space lines are not shown.

\item
Differences exist between the broad-band SED fitting procedure implemented here, and the method of \cite{Kauffmann_2003} used to estimate stellar masses in the SDSS data (see \S \ref{sec:obsdata_SDSS} and \S \ref{sec:analysis_sm_sed}). However, our methodology yields stellar masses which are more appropriate to use than the true model stellar masses when comparing to these observational results.

\item
Our SED fitting procedure uses the \cite{Calzetti_2000} dust model, whereas \cite{Davidzon_2013} also permit the Prevot-Bouchet model for VIPERS (see \S\ref{sec:obsdata_VIPERS}). However, \citeauthor{Davidzon_2013} demonstrate that using just the \citeauthor{Calzetti_2000} model has only a marginal impact on the recovered stellar mass function.

\item
As we make use of simulation snapshots in this paper (i.e.\ model outputs at fixed redshifts), rather than constructing lightcones which cover specific redshift intervals (e.g.\ \citealp{Merson_2013}), the interpretation of a sample of galaxies selected in apparent magnitude is slightly different to that for an observational survey. For example, if the model redshift is chosen to be close to the median redshift of an observational sample (as we do here), intrinsically fainter galaxies can exist in the survey sample than in the model, when the model and observations use the same faint apparent magnitude limit, due to the width of the observational redshift bin. This effect becomes important, for our purposes, for stellar masses at which the mass function becomes highly incomplete, due to the combination of redshift and apparent magnitude limits (see \S\ref{sec:analysis_sm_schfun}). In particular, this mass scale will tend to occur at higher stellar masses in a model snapshot, than in a survey sample with a lower minimum redshift than the snapshot redshift. While the use of lightcones could facilitate a more accurate treatment of the selections in apparent magnitude, the stellar mass ranges over which we compare the model clustering predictions to observational measurements are reasonably close to being complete in stellar mass. That is, we do not consider observational measurements where the survey selections result in a severely incomplete stellar mass function, and so the use of snapshots should not have a significant influence on the reliability of our comparisons. Lightcones must be used, for example, to compare theoretical model predictions to the higher redshift GAMA clustering data of Farrow et al.\ (in preparation), which we do not consider here for this reason.

\end{enumerate}

\section{Conclusions}
\label{sec:conclusions}

The stellar masses of real galaxies have to be derived from observables, typically using broad-band SED fitting. We have presented a new methodology for comparing the clustering predictions of galaxy formation models to observational data, for samples selected by stellar mass. The approach is to use estimated masses for the model galaxies, recovered using SED fitting to the model broad-band photometry. This allows us to incorporate the various systematic errors and biases involved in the fitting procedure, which can lead to significant differences with respect to the true stellar masses. These differences cannot be properly accounted for by assuming a mean rescaling, or even a rescaling plus a scatter, between the true and estimated masses, particularly for massive galaxies with strong dust attenuation \citep{Mitchell_2013}.

If stellar masses from galaxy formation models and observations can be compared in such a consistent way, the clustering of galaxies as a function of stellar mass can be used to constrain the physical processes implemented in the models, alongside traditional statistics such as the luminosity function. Our methodology demonstrates how to do such a comparison.

We have compared the clustering predictions of the Gon14 and Lac14 \textsc{galform} models to observational measurements from different surveys at redshifts of 0.1, 0.2, and 0.6. The clustering of the model galaxies as a function of stellar mass can change significantly when moving to the estimated masses. This can be understood in terms of the transfer of galaxies between stellar mass bins when estimated masses are used, which changes the distribution of halo masses in a given bin.

Considering the estimated model masses (and in particular following abundance matching to the observed stellar mass functions), we have often found that the models predict higher small-scale clustering amplitude than is inferred observationally (for projected separations $\sigma\lesssim1~h^{-1}\,\mathrm{Mpc}$). This tends to be the case for both \textsc{galform} models at the lowest redshift in our comparison (SDSS; see Fig.\ \ref{fig:wp_SDSS}c), and for the Gon14 model at higher redshifts (comparisons to GAMA and VIPERS; see Figs.\ \ref{fig:wp_GAMA}c and \ref{fig:wp_VIPERS}c). In the higher redshift comparisons, the Lac14 model tends to predict similar or lower small-scale clustering amplitude than is measured using the survey data (again see Figs.\ \ref{fig:wp_GAMA}c and \ref{fig:wp_VIPERS}c).

The small-scale clustering is sensitive to the treatment of galaxy mergers. We have introduced a new scheme for the merging of satellites with their central galaxy, in which satellite galaxies track their associated subhalo in the dark matter simulation until this is no longer resolved. In the standard \textsc{galform} approach, galaxies are assigned analytic merger time-scales as soon as they become satellites. The latter implementation yields higher small-scale clustering, as noted by \cite{Contreras_2013}, with respect to the Munich \textsc{lgalaxies} semi-analytic models. The Munich models use a merger scheme which is similar to the new scheme which we have implemented in \textsc{galform} (see \S\ref{sec:gf_sdf} for a full description of the new scheme). Using this new scheme, together with estimated stellar masses abundance matched to observationally inferred stellar mass functions, generally offers improved agreement with the observational data on small scales (particularly in the SDSS and GAMA comparisons; see Figs.\ \ref{fig:wp_SDSS}c,d and \ref{fig:wp_GAMA}c,d), or a similar overall level of agreement where the spread in the models encompasses the observational results (see the VIPERS comparison in Fig.\ \ref{fig:wp_VIPERS}c,d).

Despite reductions due to the new merger scheme, the small-scale clustering often remains higher than inferred from SDSS data, particularly for the Gon14 model (see Fig.\ \ref{fig:wp_SDSS}d). In the same comparison, the predicted clustering on the largest scales is in agreement with the SDSS measurements for only the highest masses considered. The models using the new merger scheme overall agree reasonably well with the GAMA clustering data (see Fig.\ \ref{fig:wp_GAMA}d). However, in each mass bin, only one of the two models is in good agreement with the small-scale clustering. Comparing the same models to the VIPERS measurements in Fig.\ \ref{fig:wp_VIPERS}d, the model that is closest to the observational constraints varies with stellar mass, and there can be significant differences between the modelled and observationally inferred clustering, on both small and large scales.

\cite{Kim_2009} have shown that the galaxy clustering as a function of luminosity predicted by \textsc{galform} is stronger on small scales than is measured from observations at low redshift, and thus argue for the inclusion in the models of mergers between satellite galaxies and their tidal disruption.  Such adjustments may also improve the agreement with observational measurements of the clustering as a function of stellar mass. We have not included these extensions here, but their potential implementation in future models, along with our new merger scheme, will produce galaxy formation predictions which take better account of the true dynamics of galaxies in dark matter haloes, and are as faithful as possible to the underlying dark matter distribution, allowing more robust constraints on the galaxy formation physics.

In summary, we have found that the model clustering predictions agree reasonably well with measurements from GAMA ($z=0.2$). However, we have identified some relatively large discrepancies with respect to data from SDSS ($z=0.1$) and VIPERS ($z=0.6$), considering in particular the results shown in column (d) of Figs.\ \ref{fig:wp_SDSS}, \ref{fig:wp_GAMA}, and \ref{fig:wp_VIPERS}. The differences between the model predictions and the observations vary as a function of stellar mass, redshift, and projected separation, for each \textsc{galform} model. The models considered in this paper have not been calibrated to reproduce any observational clustering data (or stellar mass functions). Our new methodology will enable future models to be calibrated to the clustering of galaxies as a function of stellar mass, thus providing more accurate and reliable constraints on how galaxies populate dark matter haloes.

\section*{Acknowledgements}

We would like to thank Sergio Contreras, Peder Norberg and Shaun Cole for useful discussions. We would also like to thank the referee, Bruno Henriques, for helpful comments and suggestions.
This work was supported by the Science and Technology Facilities Council [ST/L00075X/1].
DJRC and PDM acknowledge the support of STFC studentships [ST/K501979/1 and ST/J501013/1 respectively].
VGP acknowledges support from a European Research Council Starting Grant [DEGAS-259586].
This work used the DiRAC Data Centric system at Durham University, operated by the Institute for Computational Cosmology on behalf of the STFC DiRAC HPC Facility (www.dirac.ac.uk). This equipment was funded by BIS National E-infrastructure capital grant ST/K00042X/1, STFC capital grant ST/H008519/1, and STFC DiRAC Operations grant ST/K003267/1 and Durham University. DiRAC is part of the National E-Infrastructure.

{\footnotesize \bibliography{clustering}}

\appendix

\section{Model parameters}
\label{sec:params}

As outlined in \S\ref{sec:gf_models}, the Gon14 and Lac14 \textsc{galform} models are each based on the model of \cite{Lagos_2012}. This appendix gives details of the parameter differences between the Gon14 and Lac14 models; beyond the choice of IMF, SPS model, and default satellite merger scheme (these are described in \S\ref{sec:gf_models} and \S\ref{sec:gf_sdf}). Table \ref{tab:params} gives the values of the relevant parameters in each model. We refer the reader to \cite{Cole_2000}, \cite{Bower_2006}, and \cite{Lagos_2012} for further details of the model physics.

The total star formation rate of a galaxy $\psi$ is the sum of the contributions from quiescent star formation in the disc $\psi_\mathrm{disc}$, and from bursts $\psi_\mathrm{burst}$. The quiescent component is computed as,
\begin{equation}
\psi_\mathrm{disc} = \nu_\mathrm{SF} f_\mathrm{mol} M_\mathrm{cold,disc} ~,
\end{equation}
where $M_\mathrm{cold,disc}$ is the mass of cold gas in the disc, a fraction $f_\mathrm{mol}$ of which is molecular. $\nu_\mathrm{SF}$ is the inverse of the star formation timescale for molecular gas (see Table \ref{tab:params}).

The rate $\dot{M}_\mathrm{reheat}$ at which cold gas is reheated by supernova feedback is given by,
\begin{equation}
\dot{M}_\mathrm{reheat} = \psi \left(\frac{V_\mathrm{circ}}{V_\mathrm{hot}}\right) ^ {-\alpha_\mathrm{hot}} ~,
\end{equation}
where $V_\mathrm{circ}$ is the circular velocity of the galaxy, at the half mass radius. $V_\mathrm{hot}$ is a parameter with velocity units (see Table \ref{tab:params}), and $\alpha_\mathrm{hot}=3.2$ for both models. Gas heated by supernova feedback is assumed to be ejected beyond the virial radius, and then reincorporated into the hot halo gas reservoir at a rate proportional to $\alpha_\mathrm{reheat}$ (see Table \ref{tab:params}).

The onset of AGN feedback in \textsc{galform} requires the cooling time $t_\mathrm{cool}$ and free-fall time $t_\mathrm{ff}$ of the gas to satisfy,
\begin{equation}
t_\mathrm{cool} > t_\mathrm{ff}/\alpha_\mathrm{cool} ~,
\end{equation}
that is, the halo is assumed to be in a state of quasi-hydrostatic cooling. This criterion is controlled by the dimensionless parameter $\alpha_\mathrm{cool}$ (see Table \ref{tab:params}). Additionally, the cooling rate of the gas must be less than a fraction $\epsilon_\mathrm{SMBH}$ of the Eddington luminosity (see Table \ref{tab:params}). If both conditions are satisfied, AGN feedback is assumed to suppress the cooling of the halo gas.

The orbital energy of merging galaxies is assumed to be proportional to $f_\mathrm{orbit}$, and is used to compute the size of the spheroid formed in the merger. $f_\mathrm{burst}$ is the minimum galaxy mass (stars plus cold gas) ratio in a minor merger required to trigger bursts of star formation (major mergers occur for mass ratios above 0.3). Bursts are suppressed in minor mergers if the gas fraction in the disc of the main galaxy is less than $f_\mathrm{gasburst}$. The star formation timescale for a burst is given either by $f_\mathrm{dyn}$ times the bulge dynamical time, or by the parameter $\tau_\mathrm{min}$, whichever is larger (see Table \ref{tab:params}).

Disc instabilities, with associated bursts of star formation, are triggered when,
\begin{equation}
\frac{V_\mathrm{circ}}{\left(GM_\mathrm{disc}/r_\mathrm{disc}\right)^\frac{1}{2}}<\epsilon_\mathrm{disc}~,
\end{equation}
where $M_\mathrm{disc}$ and $r_\mathrm{disc}$ are the mass of the disc, and its half mass radius, respectively, and $\epsilon_\mathrm{disc}$ is a parameter (see Table \ref{tab:params}).

\begin{table}
\caption{
Parameters which differ between the Gon14 and Lac14 models. A description of each parameter is given in the main text.
}
\centering
\begin{tabular}{ccc}
\hline
Parameter															&Gon14			&Lac14 \\
\hline
$\nu_\mathrm{SF}~[\mathrm{Gyr^{-1}}]$		&0.5					&0.74 \\
$V_\mathrm{hot}~[\mathrm{km\,s^{-1}}]$		&425				&320 \\
$\alpha_\mathrm{reheat}$									&1.2603			&0.64 \\
$\alpha_\mathrm{cool}$										&0.6					&0.8 \\
$\epsilon_\mathrm{SMBH}$								&0.0398			&0.01 \\
$f_\mathrm{orbit}$											&1                              &0 \\
$f_\mathrm{burst}$											&0.1					&0.05 \\
$f_\mathrm{gasburst}$										&0.1					&0 \\
$f_\mathrm{dyn}$												&10					&20 \\
$\tau_\mathrm{min}~[\mathrm{Gyr}]$				&0.05				&0.1 \\
$\epsilon_\mathrm{disc}$									&0.8					&0.9 \\
\hline
\end{tabular}
\label{tab:params}
\end{table}

\label{lastpage}

\end{document}